_________________________________________________________________

\documentstyle[12pt]{article}
\newcommand{\C}{1\!\!\!C}

\newcommand{\R}{I\!\!R}
\newcommand{\Sc}{{\cal S}}
\newcommand{\be}{\begin{eqnarray*}}
\newcommand{\ee}{\end{eqnarray*}}
      \date{}
\begin{document}
\large
\setcounter{page}{1}



\title{
The  $:\!\phi^4_4\!:$ quantum field theory, II.
\\   Integrability of Wick kernels.
\footnote{This work
is supported partly by RBRF 96-0149.
It is  the third paper of the
project $\phi^4_4\cap M.$
}
}
\author{Edward  P. Osipov
\\  Department of Theoretical Physics
\\ Sobolev Institute for Mathematics
\\  630090 Novosibirsk, RUSSIA
\thanks{E-mail address: osipov@math.nsc.ru}
}
      \date{}
\maketitle
\medskip

\medskip
\begin{abstract}
We continue the construction of the
$:\!\phi^4_4\!:$ quantum field theory.
In this  paper we consider the Wick kernel of
the interacting quantum field.
Using the Fock--Bargmann--Berezin--Segal
integral representation we
prove that this kernel defines a unique
operator--valued generalized function on the space
$\Sc^\alpha(\R^4)$
 for any $\alpha<6/5,$
i.e. the constructed quantum field
is the generalized operator-valued
function of localizable
Jaffe class.
 The same assertion is valid for the outgoing
quantum field.
 These assertions about the quantum field
allow to construct the Wightman
functions, the matrix elements of the quantum scattering
operator and to consider their properties
(positivity, spectrality, Poincar\'e invariance,
locality, asymptotic  completeness, and unitarity
of the quantum scattering).

\end{abstract}


\section*{1.
 Introduction.
}

We continue the consideration of the  nonperturbative
mathematically rigorous construction of the
$:\!\phi^4_4\!:$  quantum theory in four-dimensional
space--time.

In
\cite{Osi96a}
we prove the existence and construct the complex
structure for solutions of the classical nonlinear wave
equation with the interaction $u^4_4$.
In
\cite{Osi96b}
 using this
complex structure we introduce the Wick kernel for the
interacting and outgoing quantum field and prove that these
Wick kernels define bilinear form in the Fock space of the
free incoming quantum field,
see also
\cite{Osi94a,Osi94b,Hei74,Rac75,Osi84}.

In this paper we prove that the Wick kernel introduced in
\cite{Osi96b}
defines a unique operator--valued generalized function
with test functions from $\Sc^\alpha(\R^4)$ for some
$\alpha>1$. For $\alpha>1$ the space $\Sc^\alpha(\R^4)$
contains a dense space of functions with compact support in
space--time coordinates
\cite{GelS58}.
To prove that the Wick
kernel defines an operator--valued
generalized function we
consider the Fock--Bargmann--Berezin--Segal integral
representation
\cite{Per86,Bar61,Bar67,Ber66,PanPSZ91,BaeSZ92},
 the properties of integrability
of introduced Wick kernel,
and obtain some uniform estimates.

First,  we consider the bilinear forms
$$ e^{-s_1 H}\phi(t,x)e^{-s_2 H},\quad s_1>0,\quad s_2>0, $$
and their Wick kernels.
These Wick kernels
 are defined correctly and
 define the bilinear forms.
 We prove that these Wick kernels are square integrable over
the Gaussian promeasure on the space $H^{1/2}(\R^3,\C)$ and
this square integrability is uniform for some set of
finite--dimensional complex subspaces
$L_N\subset H^{1/2}(\R^3,\C)$,
$\cup L_N = H^{1/2}(\R^3,\C)$. Moreover, we
obtain the estimate for this integral and this estimate is
uniform in $s_1$ and $s_2$. This uniform estimate and
holomorphy allow us to use the
 ``edge of the wedge" Bogoliubov theorem
\cite{Vla66,ConT79}
and to prove that the Wick kernel defines a
 unique operator--valued generalized function on some
localizable class of test functions.


\section*{2. Technical preliminaries.}

Using the paper of Paneitz et al.
\cite{PanPSZ91}
we introduce notations and (Gaussian) promeasures  that
we need.

 The following notation will be used,
unless otherwise indicated.
For a complex Hilbert space we use the inner product
$\langle ., .\rangle$ antilinear in the first argument.
  If ${\bf M}$ is a given real Hilbert space, the centered
Gaussian promeasure of covariance operator $C$ will be
denoted as $\nu_C.$
When ${\bf M}$ is
 finite--dimensional,
 $\nu_C$ is countably additive
and when ${\bf M}$ is
 infinite--dimensional,
 the promeasure $\nu_C$ is not
 in general countably additive but assigns a
unique  {\it integral} or
{\it expectation values} $E(p)$ to polynomials $p$
over ${\bf M},$ where such a polynomial is defined as a
function of the form
$p(x)=F(x_1,...,x_n),$ where $F$ is a
polynomial on $\R^n$ and $x_j=\langle e_j, x\rangle$ for
some finite set of orthonormal vectors $e_1,...,e_n$
in $D(C^{1/2}).$
We also write $E(p)=\int_{{\bf M}}\; p(x)\; d\nu_C(x)$
but note once more that the
promeasure here need
not be countably additive, so that the integral is not
necessarily of the usual Lebesgue type.
  Thus if $C=I,$
\be
E(p) &=& (2\pi)^{-n/2}\;\int_{\R^n}\; F(x_1,...,x_n)\;
 \exp[-{1\over 2}\langle x,x\rangle] dx_1...dx_n
\\ &=&
\int_{{\bf M}}\;p(x)\;d\nu_I(x).
\ee
The set of all polynomials on
${\bf M}$ will be denoted as ${\bf P}.$ The
completion of ${\bf P}$ in the $L_p$-norm,
$\| f\|_p = E(|f|^p)^{1/p},$
where $E$ and $\int_{{\bf M}}$
are extended in the obvious way to general Borel
functions of $x_1,...,x_n,$ will be denoted as
$L_p({\bf M},d\nu),$ for $p\in [1,\infty).$

As a real Hilbert space whose inner product is
$\mbox{ Re }(\langle \cdot,\cdot\rangle),$
the complex Hilbert space ${\bf M}$ will be denoted as
${\bf M^\#}.$ In ${\bf M^\#}$ we use,
unless otherwise indicated, the covariance operator
$C={1\over 2}I.$ Thus for any $p\in{\bf P(M^\#)},$ and any
orthonormal set $e_1,...,e_n$ in ${\bf M}$ such that
$p(z)=F(x_1,..., x_n, y_1,...,y_n),$ where
$\langle e_j, z\rangle = x_j+iy_j$ and
$F$ is a polynomial on
$\R^{2n},$
\be
	&&E(p)=
\\&& \pi^{-n}\;\int_{\R^{2n}}\;
 F(x_1,..., x_n, y_1,...,y_n)
\;\exp[-\| x \|^2  -
\| y\|^2]\;
dx_1...dx_n\;dy_1...dy_n,
\ee
here $\| x\|^2 = \sum x_j^2,$
 $\| y\|^2 = \sum y_j^2 .$
 The polynomials on ${\bf M^\#}$ that are holomorphic as
functions on ${\bf M}$ form a subset that will be denoted
as ${\bf P^+(M)};$ those that are antiholomorphic will be
denoted as ${\bf P^-(M)}.$ The closure in $L_p({\bf M^\#})$
(where the promeasure is understood to be $\nu_{I/2}$ unless
otherwise indicated) of ${\bf P^{\pm}(M)}$ will be denoted
as ${\bf H^{\pm}} L_p({\bf M}).$
Expectation values $E$ can be extended by
continuity to all of these spaces.

We define an entire function on a complex topological
vector space ${\bf V}$ as one whose restriction to every
finite--dimensional subspace is an entire analytic function
in the usual sense, and we denote the set of all such
functions on ${\bf V}$ as ${\bf H^+}({\bf V}).$ An
antientire function on ${\bf V}$ is defined as one whose
complex conjugate is entire, and the set of all such on
${\bf V}$ will be denoted as ${\bf H^-}({\bf V}).$ We
recall the following aspects of the representation of the
space ${\bf H}^{\pm}L_2 ({\bf M})$ as subspaces of
${\bf H}^{\pm}({\bf M})$
\cite{PanPSZ91}:

(1) If $f\in{\bf H}^{\pm}({\bf M}),$ then
$f\in{\bf H}^{\pm}L_2({\bf M})$
if and only if the supremum over all
finite--dimensional subspaces ${\bf M}$ of
$\| f|{\bf M}\|_2$
is finite, and $\| f\|_2$ is
then equal to this supremum.

(2) Let $e_1,e_2,...$ be an orthonormal basis in ${\bf M}.$
Then $F\in {\bf H}^+L_2({\bf M})$ if and only if there
exist complex numbers $a_n,$ where $n$ represents a
multiindex $(n_1,n_2,...)$ with $n_k$
 a non-negative
integer for all $k$ such that $n_k=0$ for sufficiently
large $k$ (which hereafter will be called simply
a multiindex), such that $F(z)= \sum_n a_n z^n,$ where
$$
z^n=z_1^{n_1}z_2^{n_2}...,\;\;
\quad z_j=\langle e_j, z\rangle, $$
and $\| F\|^2_2 = \sum_n\;n!\;|a_n|^2<\infty,$
where $n!=n_1!n_2!...\, .$

Let $H^\alpha(\R^3)$ (respectively,
 $H^\alpha(\R^3, \C)$) be the Hilbert space of real
valued  (respectively, complex valued) tempered
distributions such that
$(-\Delta + m^2)^{\alpha/2} f\in L_2(\R^3)$
(respectively,
$(-\Delta + m^2)^{\alpha/2} f\in L_2(\R^3,\C)$),
with scalar product
$$\langle f_1,f_2\rangle _{H^\alpha} =
\langle \overline{(-\Delta + m^2)^{\alpha/2} f_1},
(-\Delta + m^2)^{\alpha/2} f_2 \rangle _{L_2}
$$ $$= \int \overline{f_1^{\sim}(k)}(k^2+m^2)^\alpha
 f_2^{\sim}(k) d^3k,
$$
here $m>0.$ We note that the dual
space of the complex 	Hilbert space
 $H^{\alpha}(\R^3,\C)$
with the dualization
$(.,.)_{L_2}$ is the space $H^{-\alpha}(\R^3,\C).$

For the description of the complex
structure of the classical (nonlinear) wave equation
and the description of Wick kernels
we use notations from \cite{Osi96a,Osi96b}.

In \cite{Osi96b}, see also
\cite[ch.\,5.1]{BogLT69},
\cite{BaeSZ92},
\cite{PanPSZ91},
 we introduce the representation of the free
(scalar, hermitian, massive)
quantum field in the Fock space
$$\Phi o\kappa =
\oplus^\infty_{n=0} {\cal H}_n,$$
where ${\cal H}_0=\C,$
 ${\cal H}_1=
 H^{-1/2}(\R^3,\C),$
 ${\cal H}_n=
\mbox{sym}\; \widehat\otimes_n H^{-1/2}(\R^3,\C).$
The introduced Fock space
 $ {\Phi o\kappa}$
 is unitarily equivalent to the space
$${\bf H^-}(L_2(H^{1/2}(\R^3,\C)).$$
We note that the complex space
$H^{1/2}(\R^3,\C)$ is unitarily equivalent to
the space
$(H^{1/2}(\R^3), H^{-1/2}(\R^3), J)$ with the
equivalence operator $R,$ see
\cite{Osi96b}.

We introduce also spaces of test functions that we need.
 $\Sc(\R^d)$ will be denote the Schwartz space
of smooth quickly decreasing functions
on $\R^d.$
Let $\Sc^\alpha (\R^d)$ be the space of all
smooth functions on
$\R^d$ such that seminorms
$$ p_{n,a}(f)=
\sup_k(\exp(a|k|^{1/\alpha})
\vert\partial^n f^\sim(k)\vert) $$
are finite for all multiindices $n$ and
all positive $a$.
The topology on $S^\alpha$ is  given by
the seminorms $p_{n,a}$. The space $S^\alpha(\R^d)$ is
a locally convex complete countably  normed and countably
Hilbert nuclear space.
It is also a nuclear Frech{\'e}t space.
 A function from $S^1(\R^d)$ is complex analytic on some
(complex) neighbourhood of $\R^d\subset\C^n$. For
$\alpha>1$ the space $S^\alpha (\R^d)$ contains some dense
subspace of functions with compact support.
Our  spaces
$S^\alpha$ are related to the spaces of type $S$
defined in
\cite[v.2]{GelS58},
see also
\cite[v.4]{GelV61},
\cite{ConT79}.


\section*{3. Main statements.}

In
\cite{Osi96b}
we introduce the
Wick kernel
$\phi(e_{z_1},e_{z_2})$
and the
Wick symbol
$\phi(z_1,z_2)$
for the
 quantum  field with the interaction
$:\!\phi^4_4\!:,$
$$
\phi(e_{z_1},e_{z_2}) =
\exp(\langle z_1,z_2\rangle_{H^{1/2}(\R^3,\C)})
\;  \phi(z_1,z_2)
$$
$$= \exp(\langle z_1,z_2\rangle_{H^{1/2}(\R^3,\C)})
{1\over 2} (RW_{in}R^{-1}(\bar z_1+z_2)
+ \overline{RW_{in}R^{-1}
(\overline{\bar z_1+z_2})} ).
\eqno(3.1)
$$
Here we use notations introduced in
the papers \cite{Osi96a,Osi96b}.
In particular, $W_{in}$ denotes the wave
operator of the classical nonlinear equation
with the interaction $u^4_4,$
$W_{in}$ maps $in$-data into initial data
and we shall drop the subscript {\it
``in" } when not necessary.

 This Wick kernel defines
a
 unique bilinear form
\cite[Theorem 3.2]{Osi96b}
and we prove that the bilinear form is given
by a unique operator--valued generalized function on
the space
 $\Sc^1$
(and also on the Gelfand space of
 $S^1$ type),
 moreover, this
 bilinear form extends by continuity on the spaces
$\Sc^{\alpha},$ $\alpha < 6/5.$
This allows us to construct Wightman functions
(as generalized functions of Jaffe type).
 In the next papers we show that these Wightman functions
 satisfy all Wightman axioms:
 the  positivity condition,
 spectrality, Poin\-car{\'e} invariance, locality,
 and the cluster property
\cite{Osi96c}.
 A part of them is proved in this paper.
This construction gives also the possibility
to introduce the
$out$-going quantum field $\phi_{out},$ to construct
the $S$ matrix, to prove that this
operator of quantum scattering is a unitary  operator,
 and to prove that the
 asymptotic
completeness
 is fulfilled
\cite{Osi96c}.

In \cite{Osi94b} to consider a bilinear form
smoothed with test function we
considered the following bilinear form $$
\int \;e^{it_1 H}\phi(t,x)\;
e^{it_2 H} f_1(t_1)f_2(t_2)dt_1
\;dt_2,
\eqno(3.2)
$$
where $f_1,f_2$ are test functions with
 compact support in momentum space.
We proved that in this case the bilinear form
is a Wick polynomial.
 The degree of this Wick polynomial
is bounded due to the compactness of supports of
functions $f_1^\sim,f^\sim_2$ (and a strict positivity
of mass constant $m>0$).
However, holomorphy,
conservation of energy, the assertions of the type of
Reeh--Schlieder and Goodman theorems,
some natural estimates, and
{\it square integrability} of Wick kernels allow to prove
 that the smoothed bilinear form (3.2)
 is defined by a unique bounded operator for
a more extensive space of test functions,
including a dense subspace of functions with
compact support in coordinate space.

In this section we formulate the main statements
about Wick kernel (3.1) and about the bilinear form (3.2)
defined by Wick kernel smoothed with test functions
and in the next sections we shall
prove these statements.
To formulate our statements we use
notations and assertions
proved in
\cite{Osi96a,Osi96b}.
Therefore, this paper considers the proof
of following statements.

Let $ \phi_\#$
with subscript $\#$ denotes either
 the quantum $in$-coming field  $ \phi_{in},$ or
 the quantum interacting field $\phi,$
 or  the  quantum $out$-going field $\phi_{out}.$

\medskip
{\bf Theorem 3.1.}
 {\it
Let  $f_1,f_2\in\Sc ^1(\R).$
The bilinear form
$$
\int e^{it_1H}\phi_{\#}(t,h)
e^{it_2H}f_1(t_1)f_2(t_2)\;dt_1 dt_2
$$
is defined correctly on
$D_{coh}(\Sc(\R^3,\C))\times D_{coh}(\Sc(\R^3,\C)),$
 can be extended by continuity to
a unique bounded operator in the Fock space
${\Phi o\kappa}$.
This operator depends continuously
on  $f_1,f_2\in\Sc^1(\R)$
in the operator norm topology.
The following equalities are fulfilled
on
$D_{coh}(\Sc(\R^3,\C))\times D_{coh}(\Sc(\R^3,\C))$
 $$
\int \phi_\#(t+t_1,x)\;
e^{i(t_1+t_2)H}
f_1(t_1)f_2(t_2)h(x)\;dt_1 dt_2 dx
$$
$$
=\int e^{it_1H}\phi_\#(t,h)
e^{it_2H}f_1(t_1)f_2(t_2)\;dt_1 dt_2.
$$
}

\medskip
{\bf Theorem 3.2.}
 {\it
The bilinear form
$$
\int e^{i(t_1+is_1)H}
\phi_\#(t,x) e^{i(t_2+is_2)H} h(x) dx,
$$
$s_1>0,$ $s_2>0,$
is defined correctly on
$D_{coh}(\Sc(\R^3,\C))\times D_{coh}(\Sc(\R^3,\C)),$
 can be extended by continuity to
the bilinear form given by
a unique Hilbert--Schmidt operator.
This operator depends holomorphically
on  $t_1+is_1,$ $t_2+is_2$
in the strong operator topology in the Fock space.
}
\medskip

Let $V_+$ denote the  forward (light open) cone,
$$V_+=\{(s,y)\in \R^4 | \, s>|y|\}. $$

\medskip
{\bf Theorem 3.3.}
 {\it
The bilinear form
$$
e^{i(t_1+is_1)H+i(x_1+iy_1)P}
\phi_\#(t,x)
e^{i(t_2+is_2)H+i(x_2+iy_2)P}
$$
$(s_1,y_1),$ $(s_2,y_2) \in V_+,$
is defined correctly on
$$D_{coh}(\Sc(\R^3,\C))\times D_{coh}(\Sc(\R^3,\C)),$$
 can be extended by continuity to
the bilinear form given by
a unique Hilbert--Schmidt operator.
This operator depends holomorphically
on
$(t_1+is_1,x_1+iy_1),$ $(t_2+is_2,x_2+iy_2)$
(in $V_+$ and
in the norm topology of bounded operators
in the Fock space).
}

\medskip
Let $\Omega$ denotes the vacuum vector in the Fock
space of the $\phi_{in} $ field.

\medskip
{\bf Theorem 3.4.}
 {\it
Vectors
$$
e^{-s_1 H}\phi_\#(t_1,x_1)
e^{-s_2H}\phi_\#(t_2,x_2)...\Omega,
$$
$s_1>0,$ $...,$ $s_n>0,$
are defined correctly in the Fock space
${\Phi o\kappa}$
and its boundary values for
$s_1,...,s_n\to 0$ exist as vectors
in the Fock space and in the sense
 of generalized functions
 (over variables
$(t_1,x_1),..., (t_n,x_n)$)
 on any space
 $\Sc^\alpha(\R^4),$ $\alpha < 6/5,$
of test functions.
}

\medskip
{\bf Theorem 3.5.}
 {\it
Vectors
$$
e^{i(t_1+is_1)H+i(x_1+iy_1)P}
\,\phi(t'_1,x'_1)
\,e^{i(t_2+is_2)H+i(x_2+iy_2)P}
\,\phi(t'_2,x'_2)...\Omega
$$
$(s_1,y_1),...,(s_n,y_n)\in V_+,$
are defined correctly in the Fock space
${\Phi o\kappa}$
and depend complex holomorphically
on
$(t_1+is_1,x_1+iy_1),...,(t_n+is_n,x_n+iy_n).$
Its
 boundary values
 exist as vectors
in the Fock space and in the sense
 of generalized functions
 (over variables
$(t'_1,x'_1),..., (t'_n,x'_n)$)
 on any space
 $\Sc^\alpha(\R^4),$ $\alpha < 6/5,$
of test functions.
}

\medskip
{\bf Theorem 3.6.}
 {\it
The expressions
$$
(\Omega,\phi_{\#}(t_1,x_1)
...\phi_{\#}(t_n,x_n)\Omega)
$$
are defined correctly as
generalized functions
on any space
$\Sc^\alpha(\R^{4n}),$ $\alpha < 6/5.$
}

\medskip
{\bf Theorem 3.7.}
 {\it
The Wightman functions are
 generalized functions on any
  space
$\Sc^\alpha(\R^{4n}),$ $\alpha < 6/5.$
 These Wightman functions satisfy the
positivity condition,
 the spectrality condition,
  and are Poin\-car{\'e} invariant.
 The Wightman functions satisfy the
cluster property with the mass constant $m.$
}

To formulate the assertion about nontriviality we
use the notation
$$\langle e_{v_1}, \phi(t,x) e_{v_2}\rangle$$
for the value of the quantum field at a point $(t,x)$
and on coherent vectors $e_{v_1},$ $e_{v_2},$ i.e.
$$\langle e_{v_1}, \phi(t,x) e_{v_2}\rangle :=
\phi(\exp(e^{it\mu}v_1),\exp(e^{it\mu}v_2))(x).$$
With the help of this notation we formulate the
statement about nontriviality.

\medskip
{\bf Theorem 3.8} (nontriviality
\cite{Hei74,Osi94a}).
 {\it
The constructed $:\!\phi^4_4\!:$ quantum field
theory is nontrivial. The coupling constant
 $\lambda$ is determined uniquely
 by  matrix elements of the interpolating quantum field
}
\begin{eqnarray*}
\lambda
&=&{1\over 6}
\lim_{\varepsilon\to 0}\varepsilon^{-4}\lim_{T\to
\infty}
\langle e_v, e_v\rangle ^4
 (\int\langle e_v, \phi(\tau,y)
e_v\rangle ^4 d\tau dy)^{-1}
\\&&
\int (\langle e_{\varepsilon v}, \phi(t+T,x)
e_{\varepsilon v}\rangle
\langle e_{2\varepsilon v},
\dot\phi(t+T,x) e_{2\varepsilon v}\rangle
\\&&- \langle e_{2\varepsilon v}, \phi(t+T,x)
e_{2\varepsilon v}\rangle
\langle e_{\varepsilon v}, \dot\phi(t+T,x)
e_{\varepsilon v}\rangle ) dx
\\&=&{1\over 6}\lim_{\varepsilon\to 0}
\varepsilon^{-4}
\langle e_v,e_v\rangle ^4
(\int\langle e_v,\phi(\tau,y)
e_v\rangle ^4 d\tau dy)^{-1}
\\&& \int(\langle e_{\varepsilon v},
\phi_{out}(t,x) e_{\varepsilon v}\rangle
 \langle e_{2\varepsilon v},
 \dot\phi_{out}(t,x) e_{2\varepsilon v}\rangle
\\&&- \langle e_{2\varepsilon v},
 \phi_{out}(t,x) e_{2\varepsilon v}\rangle
 \langle e_{\varepsilon v}, \dot\phi_{out}(t,x)
e_{\varepsilon v}\rangle ) dx
 \end{eqnarray*} {\it for any}
 $v\in \Sc(\R^3,\C),$ $v\not=0.$

\medskip


\section*{4. Integrability of Wick kernels.}

In this section we consider the introduced
Wick kernel (3.1) and properties of its
integrability. First, we formulate  the assertion
about holomorphy.

\medskip
{\bf Theorem 4.1.}
 {\it
Let
$(z_1,z_2) \to \phi(e_{z_1},e_{z_2}),$
$(z_1,z_2) \to \phi_{out}(e_{z_1},e_{z_2}),$
$$ \phi(e_{z_1},e_{z_2}) =
\exp(\langle z_1,z_2\rangle_{H^{1/2}(\R^3,\C)})
 {1\over 2}
(RWR^{-1}(\bar z_1+ z_2)
 + \overline{RWR^{-1}(\overline{\bar z_1+ z_2})}),
$$
$$
\phi_{out}(e_{z_1},e_{z_2})
=\exp(\langle z_1,z_2\rangle_{H^{1/2}(\R,\C)})
{1\over 2}(RSR^{-1}(\bar z_1+ z_2) +
\overline{RSR^{-1}(\overline{\bar z_1 + z_2})}),
$$
be Wick kernels defined on
$H^{1}(\R^3,\C) \times H^{1}(\R^3,\C)$
 $(\subset
H^{1/2}(\R^3,\C) \times H^{1/2}(\R^3,\C)).$
 Then
for $(z_1,z_2)\in H^{1}(\R^3,\C)$ \,
$\phi_\#(e_{z_1},e_{z_2}) \in
 H^{1}(\R^3,\C)$
and
$$\phi_\#(e(e^{i(t_1+is_1)\mu+i(x_1+iy_1)p}z_1),
e(e^{i(t_2+is_2)\mu+i(x_2+iy_2)p}z_2))
$$
are functions antiholomorphic on $(t_1+is_1,x_1+iy_1)$ and
 holomorphic on
$(t_2+is_2,x_2+iy_2)$ in $(\R^4 + iV_+,
\R^4 + iV_+)$
with values in $H^1(\R^3,\C)$ $\subset$
$H^{1/2}(\R^3,\C).$
Here
$\mu=(-\Delta+m^2)^{1/2},$
$p = -i \partial/\partial x = -igrad,$
$ \phi_\#$
is either the in-coming field $\phi_{in},$
or the interacting field $\phi,$
or the out-going field $\phi_{out}.$
}

\medskip
{\it Proof of Theorem 4.1.}
  Theorem 4.1
is  a direct consequence of
 the solution properties of  the classical nonlinear
wave equation.
This theorem is implied by the assertion
about holomorphy, proved in
\cite[Theorem 1.1]{Osi96a},
and by holomorphic dependence on
$(t+is, x+iy).$ The transform
$$
e^{i(t+is)\mu+i(x+iy)(-igrad)}
$$
depends holomorphic on
  $(t+is, x+iy) \in \R^4\oplus iV_+$
and acts in
$H^{1}(\R^3,\C)$
 $\subset H^{1/2}(\R^3,\C).$
 Theorem 4.1 is proved.

\medskip
We show now that the Wick kernel of the
 interacting field corresponding to the expression
$$\int e^{-s_1H}\phi(t,x)e^{-s_2H} h(x) dx$$
is in
terms of the paper \cite{PanPSZ91} a
 standard effectively
square integrable analytic kernel with respect to
the dynamics and the expression is given by
 a unique Hil\-bert-\-Schmidt operator
in the Fock space.

We prove the following assertions.

\medskip
{\bf Theorem 4.2.}
 {\it
Let
$$
\phi(e_{z_1},e_{z_2})
=\exp(\langle{z_1,z_2}\rangle_{H^{1/2}(\R^3,\C)})
{1\over 2}
(RWR^{-1}(\bar z_1 + z_2)
 + \overline{RWR^{-1}(\overline{\bar z_1 + z_2})})
$$
be the Wick kernel.
This Wick kernel as a function
$$
(z_1,z_2)\to\phi(e_{z_1},e_{z_2})
$$
is the  function on
$
 H^1(\R^3,\C)  \times
 H^1(\R^3,\C)$
 with values in
$ H^1(\R^3,\C),$  antiholomorphic on
$z_1$ and holomorphic on $z_2.$

This Wick kernel
 gives correctly defined bilinear form (in the Fock
Hil-\-bert space).  The expression $$
\phi(t_1+is_1,t_2+is_2,e_{z_1},e_{z_2})
$$ $$:=
 \int(\phi(e(e^{it_1\mu-s_1\mu}z_1),
e(e^{it_2\mu-s_2\mu}z_2))(x)\,
h(x) \,dx,
$$
$h(x) \in \Sc(\R^3),$ $s_1, s_2\ge 0,$
considered as a Wick kernel, is
defined correctly.
For $s_1, s_2 > 0$
this  Wick kernel is
square integrable over the Gaussian promeasure on
some sequence of finite-\-dimensional subspaces
$L_N\oplus L_N,$ $L_N\subset \Sc(\R^3,\C),$ $\cup_N L_N$
is dense
in $H^{1}(\R^3,\C)$ and in $H^{1/2}(\R^3,\C).$

The uniform estimate on $t_1,$ $s_1,$ $t_2,$
$s_2$ is valid,
$$
\sup_{L_{N_1},L_{N_2}}
\int_{L_{N_1}}
\int_{L_{N_2}}
 |\phi(t_1+is_1,t_2+is_2,e_{z_1},e_{z_2})|^2
 d\nu(z_1) d\nu(z_2)
$$
$$\leq c_1\exp(c s_1^{-2\alpha-1}+c s_2^{-2\alpha-1})
\;\
 \Vert\mu^{-1/2}h\Vert^2,
$$
 $\alpha$ is such that
 $\Vert\mu^{-\alpha}\Lambda(\Sigma,\delta)
\Vert_{HS(H^{1/2}(\R^3))} <\infty,$
for instance, $\alpha>2$ (see Appendix).
 $\Lambda(\Sigma,\delta)$ is the localization operator
(see Section 5)
constructed for the causal shadow
$\Sigma({\cal O})$ at time zero of some open bounded set
${\cal O} \subset \R^4.$

Therefore,
$\phi(t_1+is_1,t_2+is_2,e_{z_1},e_{z_2})$
is a standard kernel in sense of
\cite{PanPSZ91} belonging to
$L_2(H^{1/2}(\R^3,\C) \oplus H^{1/2}(\R^3,\C)).$
 There exists a unique Hil\-bert-\-Schmidt
operator with the Wick  kernel coinciding with the Wick
 kernel
$\phi(t_1+is_1,t_2+is_2,e_{z_1},e_{z_2})$
 on
$  H^1(\R^3,\C)  \times  H^1(\R^3,\C).$

The same assertion is valid for the Wick kernel
of  the $out$-going field
$$
\phi_{out}(e_{z_1},e_{z_2})
= \exp({\langle z_1,z_2\rangle}_{H^{1/2}(\R^3,\C)}
)
{1\over 2}(RSR^{-1}(\bar z_1+ z_2) +
\overline{RSR^{-1}(\overline{\bar z_1+ z_2})}).
$$
}
\medskip

{\bf Remark.}
In particular,
in terms of Paneitz, Pedersen, Segal,
Zhou \cite{PanPSZ91}
Theorem 4.2 means that the Wick kernel of
a smoothed  interacting quantum field
of the $:\!\phi^4_4\!:$ theory  is
a standard kernel on the complex Hilbert space
$H^1(\R^3,\C)$ and with respect to the dynamics
described
in coordinates of the $in$-field
(Paneitz et al. \cite{PanPSZ91}
use an ``old" physical term, so called single-\-particle
Hamiltonian).

\medskip
{\bf Theorem 4.3.}
 {\it
Let
$(t_1+is_1,x_1+iy_1), (t_2+is_2,x_2+iy_2) \in \R^4+iV_+,$
i.e., $s_1>|y_1|,$ $s_2>|y_2|.$

The expression
$$
\phi(t_1+is_1,x_1+iy_1,t_2+is_2,x_2+iy_2,e_{z_1},e_{z_2})
$$ $$:=
 \int(\phi(e(e^{(it_1-s_1)\mu+(x_1-iy_1)grad}z_1),
e(e^{(it_2-s_2)\mu+(x_2-iy_2)grad}z_2))(x)\,
h(x) \,dx,
\eqno(4.1)$$
$h(x) \in \Sc(\R^3),$
considered as a Wick kernel, is
defined correctly.
For $s_1>|y_1|, s_2 > |y_2|$
this  Wick kernel is
square integrable over the Gaussian promeasure on
some sequence of finite-\-dimensional subspaces
$L_N\oplus L_N,$ $L_N\subset \Sc(\R^3,\C),$ $\cup_N L_N$
is dense
in $H^{1}(\R^3,\C)$ and in $H^{1/2}(\R^3,\C).$

The uniform estimate on $t_j,$
$s_j,$ $x_j,$
 $y_j,$
$j=1,2,$ is valid
$$
\sup_{L_{N_1},L_{N_2}}
\int_{L_{N_1}}
\int_{L_{N_2}}
 |\phi(t_1+is_1,x_1+iy_1,t_2+is_2,x_2+iy_2,
 e_{z_1},e_{z_2})|^2 d\nu(z_1) d\nu(z_2)
$$
$$ \leq c_1\exp(c (s_1-|y_1|)^{-2\alpha-1}
+c (s_2-|y_2|)^{-2\alpha-1})
\;\
 \Vert\mu^{-1/2}h\Vert^2,
$$
 $\alpha$ is such that
 $\Vert\mu^{-\alpha}\Lambda(\Sigma,\delta)
\Vert_{HS(H^{1/2}(\R^3))} <\infty,$
for instance, $\alpha>2$ (see Appendix).
 $\Lambda(\Sigma,\delta)$ is the localization operator
(see Section 5)
constructed for the causal shadow
$\Sigma({\cal O})$ at time zero of some open bounded set
${\cal O} \subset \R^4.$

Therefore,
$\phi(t_1+is_1,x_1+iy_1,
t_2+is_2,x_2+iy_2,e_{z_1},e_{z_2})$
is a standard kernel in sense of
\cite{PanPSZ91} belonging to
$L_2(H^{1/2}(\R^3,\C) \oplus H^{1/2}(\R^3,\C)).$
 There exists a unique Hil\-bert-\-Schmidt
 operator
(in the Fock Hilbert space)  with the Wick  kernel
 coinciding with
(4.1)
  on $ H^1(\R^3,\C)
\times  H^1(\R^3,\C).$

The same assertion is valid for the Wick kernel
of  the $out$-going field
$$
\phi_{out}(e_{z_1},e_{z_2})
= \exp({\langle z_1,z_2\rangle}_{H^{1/2}(\R^3,\C)}
)
{1\over 2}(RSR^{-1}(\bar z_1+ z_2) +
\overline{RSR^{-1}(\overline{\bar z_1+ z_2})}).
$$
}

\medskip
{\bf Remark.}
The Wick kernel
$\phi(t_1+is_1,x_1+iy_1,
t_2+is_2,x_2+iy_2,e_{z_1},e_{z_2})$
corresponds to the expression
$$
\int e^{(it_1-s_1)H+(ix_1-y_1)P}
\phi (0,x)
 e^{(it_2-s_2)H+(ix_2-y_2)P}
h(x) dx
$$
on the coherent vectors
$e_{z_1},$ $e_{z_2}.$

\medskip
{\it Proof of Theorem  4.2 and Theorem  4.3.}
 We consider the expression
$$
\phi(e_{z_1},e_{z_2})=\exp({\langle z_1,z_2\rangle})
{1\over 2}
(RWR^{-1}(\bar z_1 +  z_2)
 + \overline{RWR^{-1}(\overline{\bar z_1 + z_2})}).
$$
This expression is
 defined correctly for $z_1,$ $z_2$ $\in$
$ H^{1}(\R^3,\C)$ and is holomorphic on
$\bar z_1, z_2$ from
 $H^1(\R^3,\C)$ into $H^1(\R^3,\C)$
\cite[Theorem 1.1]{Osi96a}.
Thus, this expression defines
a Wick kernel.

We consider now the kernel
$$
\phi(t_1+is_1,x_1+iy_1, t_2+is_2,x_2+iy_2,e_{z_1},e_{z_2})
$$
and  prove that
this kernel defines a unique
Hil\-bert-\-Schmidt operator. To prove this
assertion we show that the holomorphic kernel is
square integrable uniformly with respect
to  a dense sequence of finite--dimensional
subspaces in the (anti)holomorphic representation by
Paneitz, Pedersen, Segal, Zhou
\cite{PanPSZ91}
for the $in$--field.

We need to obtain the uniform estimate for
integrals of squared absolute value of
 Wick kernels of
fields $\phi_\#$ over dense sequence of
finite-\-dimensional complex linear
subspaces  $L_N$ from
$H^{1/2}(\R^3,\C).$

To obtain this estimate we introduce localization domains
and construct dense sequence of complex linear
finite-\-dimensional subspaces $L_N\subset
H^{1/2}(\R^3,\C).$
We construct this sequence of finite-dimensional
subspaces with the help of a basis consisting of
real and localized vectors and use also operators
of localization. Properties of these operators of
localization are considered in Section 5.

To construct (complex linear)
fi\-ni\-te-\-di\-men\-sion\-al subspaces in
the complex Hilbert space $H^{1/2}(\R^3,\C)$
we introduce some bounded open set
${\cal O}\subset\R^4$ and a subspace
$$
\Sc_{Re}({\cal O})\subset \Sc(\R^4),
$$
consisting of real-valued  functions with support in
${\cal O}.$ We introduce also subspaces
$$
\Sc_{Re}^{even}({\cal O})=
\{f\in\Sc_{Re}({\cal O})\vert f^\Theta=f\},
$$
 $$
\Sc_{Re}^{odd}({\cal O})=
\{f\in\Sc_{Re}({\cal O})\vert f^\Theta=-f\},
$$
here  $f^\Theta(t,x)=f(-t,x).$ We choose
${\cal O}$ such, that ${\cal O}^\Theta={\cal O},$
i.e.,
 we choose  a set
${\cal O}$ and a time reflection, such
that ${\cal O}$ would be symmetric with respect to
 the time reflection.  Here
$$
{\cal O}^\Theta
=\{(t,x)\in\R^4\, \vert \,(-t,x)\in {\cal O}\}.
$$

We remark, first of all, that our choice of ${\cal O}$
(such that ${\cal O}^\Theta = {\cal O}$)
 satisfies the following conditions:
$$
\Sc_{Re}^{even}({\cal O})\subset \Sc_{Re}({\cal O}),
\quad
\Sc_{Re}^{odd}({\cal O})\subset \Sc_{Re}({\cal O}),
$$
 if
 $f\in\Sc_{Re}^{even}({\cal O}),$
then $\mbox{ Re }Pf = Pf,$
and if  $f\in \Sc_{Re}^{odd}({\cal O}),$
then  $i\mbox{ Im }Pf = Pf.$
The projection operator $P$ is defined in Sect. 5.

Further,
$$
\Sc_{Re}({\cal O})=\Sc_{Re}^{even}({\cal O})\oplus
\Sc_{Re}^{odd}({\cal O}).
$$
In this direct sum  (with
a choice ${\cal O}^\Theta = {\cal O}$)
vectors of the spaces
$P\Sc_{Re}^{even}({\cal O})$ and
$P\Sc_{Re}^{odd}({\cal O})$
  are vectors given by pure real
and, respectively, by pure
imaginary functions from the Hilbert space
 $H^{1/2}(\R^3,\C),$
 or they are pure real and
pure imaginary  vectors in the
 standard realification
 $$(H^{1/2}(\R^3) \oplus i H^{1/2}(\R^3),
\langle.,.\rangle_{H^{1/2}(\R^3,\C)}),$$
or pure real and
pure imaginary  vectors in the
 standard realification
 $$(H^{1/2}(\R^3), H^{-1/2}(\R^3))$$
 of the complex Hilbert space
 $(H^{1/2}(\R^3), H^{-1/2}(\R^3), J)$
with the operator of imaginary unit $J$ and
given by the unitary isomorphism $R,$
 $R(\varphi,\pi)= \varphi +i\mu^{-1}\pi. $

Thus,
$$P\Sc_{Re}^{even}({\cal O})
= \mbox{Re }P\Sc_{Re}({\cal O}).$$
A density of  the space
$P\Sc_{Re}({\cal O})$ in
 $H^{1/2}(\R^3,\C)$ (see, Lemma 5.1 in Section 5) implies that
$P\Sc_{Re}^{even}({\cal O})$
 is dense in the space
  $H^{1/2}(\R^3)$
as a real subspace
 $H^{1/2}(\R^3,\C)$
(and considered as a complex linear space it
 is dense in
 $H^{1/2}(\R^3,\C)$ ).
The definition of the projection
operator $P$ on the mass hyperboloid
(in momentum space)  for a function of four-\-dimensional
argument  (in coordinate space)
is given in  Section 5.
The projection operator $P$
 coincides, in fact, with the
projection operator of Goodman
\cite{Goo64},
see also \cite[Introduction]{Osi96a}.

To introduce a basis of finite dimensional complex
linear subspace $L_N \subset H^{1/2}(\R^3,\C)$
with the help of real and localized vectors we consider
real-linearly independent vectors in $H^{1/2}(\R^3)$ of
the form
$e_j \in P\Sc_{Re}^{even}({\cal O}),$
$j = 1, ..., N.$ It is easy to see that these vectors exist.
 (This existence follows from the inclusion
$P\Sc_{Re}^{even}({\cal O}) \subset H^{1/2}(\R^3)$
and a density of
$P\Sc_{Re}^{even}({\cal O})$ in $H^{1/2}(\R^3),$
see Section 5).
Since
the considered vectors are real
the linear transformations
defined the orthonormalized (real)
finite-\-dimensional basis in
 $H^{1/2}(\R^3) ( \subset
H^{1/2}(\R^3,\C))$
are given by real matrices. It is easy to see
that in this case the basis vectors belong to
 $\Sc_{Re}(\R^3),$ this is a consequence of
strictly positivity of the
mass constant.

We denote these vectors as
$e^{Re}_j,$ $j=1,...,N.$
Finally, the complex linear
envelope of vectors  $\{e^{Re}_j, j=1,...,N\}$
defines the complex linear finite-\-dimensional
subspace
$L_N\subset H^{1/2}(\R^3,\C),$  that we need.
In this case the vectors
$\{e^{Re}_j, j = 1,...,N\}$
define the orthonormalized basis in
$L_N,$
$e^{Re}_j\in P\Sc_{Re}^{even}({\cal O}),$ $j = 1,...,N,$
$\mbox{ Re }e^{Re}_j=e^{Re}_j,$
$$
\langle e^{Re}_j,e^{Re}_{j'}\rangle_{H^{1/2}(\R^3,\C)}
= \langle e^{Re}_j, e^{Re}_{j'}\rangle_{H^{1/2}(\R^3)}
=\delta_{jj'}.
$$

Moreover, the complex linear envelope of vectors
$\{e^{Re}_j, j = 1,...,N\}$
generates the whole complex linear space
 $L_N.$ It is isomorphic to $\C^N$
in the chosen basis. The real linear envelope of vectors of
basis  $\{e^{Re}_j, j = 1,...,N\}$, or, correspondingly,
of vectors $\{ie^{Re}_j, j = 1,...,N\}$,  generates real
linear spaces isomorphic to  $\R^N$ and corresponding
to the pure real and pure imaginary subspaces of $\C^N.$

We remark that although vectors
$ie^{Re}_j\in H^{1/2}(\R^3,\C),$
but as a nonzero vector
$ie^{Re}_j$  $\not\in
P\Sc_{Re}^{even}({\cal O}).$
 We emphasize that the map  $P$ is real linear only and,
in general,  $i P\Sc_{Re}({\cal O})\not\subset
 P\Sc_{Re}({\cal O}).$ However one can easily see that
$$
Pf^\Theta=\overline{Pf},\quad f\in \Sc_{Re}({\cal O}).
$$
 Here  (and in the previous exposition) the
notations ``Re'', ``Im'' are used both to denote
subscript
and superscript
and to denote the standard real and,
respectively,
imaginary
part of a vector $z,$ considered as a
complex-\-valued function
$$
z=\mbox{ Re }z+i\mbox{ Im }z.
$$
We remark that our choice of the basis
$\{e^{Re}_j, j=1,...,N\}$ in $\C^N$ and $\{e^{Re}_j,
ie^{Re}_j, j=1,...,N\}$
in $\R^N \oplus i\R^N$ gives that the determinant of this
transformation is   $1/2^N.$ This is in agreement with the choice
of the Gaussian promeasure
in complex variables with covariance
  $I/2$  and in real variables
with covariance
 $I,$ see
 \cite{PanPSZ91}.

For a (bounded open) set
${\cal O} \subset \R^4$
let $\Sigma({\cal O})$ denote its
causal envelope  on the hyperplane
$t=0,$
$$
\Sigma({\cal O})
= \{x \in \R^3 \;\vert\;\exists (t,y)
\in {\cal O}, \quad t^2-(x-y)^2\geq 0\}. $$
 We
 use the localization operators for  ${\cal O}$
introduced in Section 5,
in particular,
 the complex linear operator
$\Lambda(\Sigma,\delta) =$ $\mu^{-1}\Lambda\mu =$
$R\Lambda_{\C}(\Sigma,\delta)R^{-1},$
$$
\Lambda_{\C}(\Sigma,\delta)=
\pmatrix{\mu^{-1}\Lambda\mu&0\cr 0&\Lambda\cr},
$$
where $\Lambda$ is
the operator of multiplication
on a smooth function $\Lambda(x)$ in coordinate space
 equal $1$ on
$\Sigma({\cal O})$ and having compact support.
 With our choice of the basis
$\mu^{-1}\Lambda\mu \,e^{Re}_j=e^{Re}_j$
and for vectors $z$
from  $L_N\;\;$
  $z=$ $\mu^{-1}\Lambda\mu z =$
$\Lambda(\Sigma,\delta) z=$
$R\Lambda_{\C}(\Sigma,\delta)R^{-1} z .$

\medskip
{\bf Remarks.}
 1. The choice of basis with the help of real and
appropriate functions (i.e. with the help of real,
smooth functions with
compact support in space-\-time) allows to introduce
an operator of localization consistent with
the complex structure.

2. We consider here the simplest
variant of the choice of localization
region  ${\cal O}$ and (time) reflection.
It is possible to take the time reflection not necessary
with respect to the zero moment. Indeed if
we have a region of localization   ${\cal O}$
and an operator
of localization
$\Lambda(\Sigma,\delta)$
then we can choose some smaller region of localization
 ${\cal O}_1\subset{\cal O}$
and the reflection  $\Theta(a_0)$ with respect
to a time moment  $a_0,$ $(a_0,a)\in{\cal O}_1,$ such,
that
${\cal O}_1^{\Theta(a_0)}={\cal O}_1.$

3. The consideration of
$PT$ reflection (more precisely
 $CPT$ reflection)
allows to define physically adequate the
``real'' and ``imaginary'' parts of complex spaces.

4. We emphasize that $\mu^{-1}\Lambda\mu$
is a complex linear operator. It commutes with the
 imaginary unit $i\equiv\sqrt{-1}.$
 The operator  $\Lambda_{\C}(\Sigma,\delta)$
 is connected via the isomorphism $R$
 with the operator
  $\mu^{-1}\Lambda\mu,$
   $\Lambda_{\C}(\Sigma,\delta)$ $=$
$R^{-1}\mu^{-1}\Lambda\mu R,$
and the operator  $\Lambda_{\C}(\Sigma,\delta)$
commutes with the operator of imaginary unit
$J,$
 $J=R^{-1}iR.$

5. In addition,
 the transformation
$$
P:\Sc_{Re}({\cal O})\to H^{1/2}(\R^3,\C),
$$
  $$
(Pf)(x) = \int(\mu^{-1}\cos(\mu t))(x-y) f(t,y)dtdy
$$
$$ +i\int(\mu^{-1} \sin(\mu t))(x-y) f(t,y)dtdy,
$$
is a real linear operator only.
It is important point for
the next presentation. It is connected with a possibility
to choose a real basis in
 the subspace $P\Sc_{Re}({\cal O}),$
which is dense in  $H^{1/2}(\R^3,\C).$
 The $T$--symmetry (or the more general
case of $CPT$--symmetry) is connected
 namely with this point.
 In the case of considered interaction
  $T$--symmetry fulfilled certainly. In the
classical case it means
that if $u(t,x)$  is a solution of the nonlinear
equation, then  $u(-t,x)$ is also a solution of
this nonlinear equation). This symmetry connects
nonlinearity and holomorphy for the case of system with
infinite degrees of freedom
and appears as the fundamental quantum
$CPT$--symmetry.
 The subspace  $
P\Sc_{Re}({\cal O})
$
is a dense subspace in  $D(\mu^{1/2}),$
as in the domain of definition of
the operator  $\mu^{1/2}$ with
appropriate topology in
$H^{1/2}(\R^3,\C)$) and, therefore, it is dense in
$H^{1/2}(\R^3,\C).$
 However, $P$ is
 a real linear map,
 in general,
$iP\Sc_{Re}({\cal O})
\not\subset P\Sc_{Re}({\cal O})),$
but, of course,
$iP\Sc_{Re}({\cal O})
\subset H^{1/2}(\R^3,\C).$

\medskip
\medskip
\medskip
Further, for $z\in P\Sc_{Re}({\cal O})$
$
z=R\Lambda_{Re}(\Sigma,\delta)R^{-1}z,
$
 for
 $z\in P\Sc_{Re}^{even}({\cal O})$
$$
z=\Lambda(\Sigma,\delta)z
=\mu^{-1}\Lambda\mu z =
R\Lambda_{\C}(\Sigma,\delta)R^{-1}z
=R\Lambda_{Re}(\Sigma,\delta)R^{-1}z,
$$ and for $z\in L_N$
$$ z=\Lambda(\Sigma,\delta)z
=\mu^{-1}\Lambda\mu z =
R\Lambda_{\C}(\Sigma,\delta)R^{-1}z.
$$
In fact, if
$z=Pf,$ $f\in \Sc_{Re}({\cal O}),$
then
$$z=\mu^{-1}\int\cos\mu tf(t)dt
+i\mu^{-1}\int\sin\mu t f(t)dt,$$
$$
R\Lambda_{Re}(\Sigma,\delta)R^{-1}z
=R\pmatrix{\mu^{-1}\Lambda\mu,&0\cr
0,&\mu\Lambda\mu^{-1}\cr}R^{-1} Pf
$$
$$
=R\pmatrix{\mu^{-1}\Lambda\mu,&0\cr
0,&\mu\Lambda\mu^{-1}\cr}R^{-1}
(\mu^{-1}\int\cos\mu tf(t)dt
+ i\mu^{-1}\int\sin\mu t f(t)dt)
$$
$$
=R\pmatrix{\mu^{-1}\Lambda\mu,&0\cr
0,&\mu\Lambda\mu^{-1}\cr}
(\mu^{-1}\int\cos\mu tf(t)dt,
+\int\sin\mu t f(t)dt)
$$
$$
=R(\mu^{-1}\Lambda\int\cos\mu tf(t)dt,
+\mu\Lambda\mu^{-1}
\int\sin\mu t f(t)dt)
$$
$$
=R(\mu^{-1}\int\cos\mu tf(t)dt,
+\int\sin\mu t f(t)dt)
$$
$$
=\mu^{-1}\int\cos\mu tf(t)dt
+i\mu^{-1}\int\sin\mu t f(t)dt=Pf=z.
$$
Since the basis
$e^{Re}_j$ $\in$ $P\Sc^{even}_{Re}({\cal O}),$
the other equalities can be verified analogously.

Thus, with the help of a localization operator
$\Lambda(\Sigma,\delta),$
introduced in Section 5, we can write the
following equality
$$
\phi(e(A_1 z_1),e(A_2 z_2))
=\phi(e(A_1\Lambda(\Sigma,\delta)z_1),
e(A_2\Lambda(\Sigma,\delta)z_2),
\eqno(4.2)
$$
here $A_1,$ $A_2$ are the operators defined by
the dynamics.
This equality is fulfilled on
the sequence of complex linear
 finite-\-dimensional subspaces $L_N,$
which
is dense in
 $H^1(\R^3,\C).$
Eq. (4.2)
 is implied by Theorem 5.2,
or Corollary 5.3.

This gives us the possibility
to consider the expression
$$
\phi(e(\exp(-s_1\mu)\Lambda(\Sigma,\delta)z_1),
e(\exp(-s_2\mu)\Lambda(\Sigma,\delta)z_2)),
$$
as the Wick kernel
defined on the subspaces
 which are  dense in
$H^1(\R^3,\C)$ and in $H^{1/2}(\R^3,\C).$
 We can also use these equalities to obtain
the required estimates.

The considered
kernel is defined with the help of an
operator of localization,
but it defines the same  operator, that
corresponding to the bilinear form
$$
\int e^{-s_1 H} e^{it_1 H}
\phi(0,x) e^{it_2 H} e^{-s_2 H}
h(x)dx
$$
and, therefore, the obtained estimate depends
on localization and is independent of dimensions
of finite-\-dimensional subspaces.

More accurately, the independence of localization
is the consequence of Lemma 1.1,
 Lemma 1.5
\cite[Lemma 1.1, Lemma 1.5]{PanPSZ91}
and the fact, that the operator is defined
by its Wick kernel and integration over promeasure,
i.e., over compatible sequence of measures on
 finite-\-dimensional subspaces whose union is dense.

As a result,  localization and complex structure
imply a
 uniform square integrability of Wick kernels
(taken at ``imaginary times''). This is the consequence
of the possibility to coordinate the complex structure
with
(fundamental from physical point of view) symmetries
and properties.
The
 complex structure
 and the integrability are build into
 the foundation
 and the integrability of Wick kernel is
 based on a complex structure, causality,
translation invariance, and spectrality.

Here the fundamental importance is that the considered
complex Hilbert space is a pair of real Hilbert
spaces with
the (physically natural)
 realification.
This
 realification
 is, in fact, initial data,
 the canonical coordinate
and the  canonical momentum.

Uniform (for square integrability of Wick kernel)
 estimates over
 some dense sequence of
finite-\-dimensional subspaces with standard realification
 and holomorphy allow to show that
 the
 considered Wick kernel defines a unique
Hil\-bert-\-Schmidt operator in the Fock space.

To prove the Wick kernel
$K(e_{z_1},e_{z_2})$ is effectively standard
(in definition of the paper
\cite{PanPSZ91})
and defines  a (unique)
 Hil\-bert-\-Schmidt
 operator in the Fock space we use
uniform estimates, holomorphy
(Theorem 2.1 \cite{Osi96b}, or Theorem 1.1
\cite{Osi96a}), and apply Lemma 1.1
 \cite{PanPSZ91} and
Lemma 1.5
\cite{PanPSZ91}.

Now turn to the details  and obtaining the estimates.

We introduce the following Wick kernels
\be
K(e_{z_1},e_{z_2})
&=&
\exp(\langle A_1 z_1, A_2 z_2
\rangle_{H^{1/2}(\R^3,\C)})
\cr &&\langle h,
 RWR^{-1}( \overline{A_1 z_1} + A_2 z_2)
\rangle_{H^{1/2}(\R^3,\C)},
\cr
K_{out}(e_{z_1},e_{z_2})
&=&
\exp(\langle A_1 z_1, A_2 z_2
\rangle_{H^{1/2}(\R^3,\C)})
\cr &&\langle h,
 RSR^{-1}( \overline{A_1 z_1} + A_2 z_2)
\rangle_{H^{1/2}(\R^3,\C)},
 \ee
 where the operators $A_1,$ $A_2$
are the operators defined on the complex Hilbert space
$H^{1/2}(\R^3,\C),$
\be
(A_j z_j)^\sim(k)
&=&
 e^{-s_j\mu(k) -y_j k + it_j\mu(k)+ix_j k}
z^\sim_j(k),\quad j=1,2.
\ee

We introduce also the operators
$$
T_j z= e^{-(s_j-|y_j|)\mu}
\Lambda(\Sigma, \delta), \quad j=1, 2.
$$
Here
$\Lambda(\Sigma, \delta)$ is a localization operator
 corresponding to the chosen (open bounded) set
 $\cal O$ $\subset$ $\R^4,$ see Section 5.
 We remark that
$$\Lambda(\Sigma, \delta)
=\mu^{-1}\Lambda\mu
=R\Lambda_{\C} (\Sigma, \delta)R^{-1}
 $$
 and on $L_N$
 and on the complex linear span of
$P\Sc^{even}_{Re}({\cal O})$
$$
T_j z = e^{-(s_j-|y_j|)\mu}z
 \eqno(4.3)
 $$
 (see Theorem 5.2 and Corollary 5.3).
Here $\Lambda$ is the operator of multiplication
on a function in coordinate space equal $1$ on
$\Sigma({\cal O}).$

It should be noted that localization operators
appear namely in the consideration
of finite-\-dimensional subspaces in
$H^{1/2}(\R^3,\C)$ and in estimates.
 Since  Wick kernels are defined and
holomorphic on all coherent vectors
with finite energies
the field as operator-\-valued function
does not depend, in fact,
on domain of localization and
on operator of localization. This is the consequence
of the equality (4.2)
and
 the consequence of
 the density of the space
 $P\Sc_{Re}({\cal O})$
 in $H^{1/2}(\R^3,\C)$
for any open sets ${\cal O}.$
This is also a direct consequence of the principle
of causality
(in the form of hyperbolicity of the wave equation
and/or in the form of the finite propagation
velocity).

Thus, a uniform square integrability of the Wick kernel
over fi\-ni\-te-\-di\-men\-sion\-al subspaces follows,
in fact, from  energy conservation and
causality.
Due to the energy conservation and
the causality principle
 ($+$ {\it ho\-lo\-morphy!}) it is sufficient
 to consider the
Wick kernel on the
subspace
which is the complex linear span of
$P\Sc_{Re}^{even}({\cal O}).$

\medskip
\medskip
Let us consider first the case of the
kernel  $K(e_{z_1},e_{z_2})$  with
$(s_1,y_1) = (s_1,0),$ $(s_2,y_2) = (s_2,0).$

The kernel
$K(e_{z_1},e_{z_2})$ is antiholomorphic in
 $z_1,$ holomorphic in
 $z_2$ (see
Theorem 2.1 in  \cite{Osi96b}
and
Theorem 1.1 in \cite{Osi96a})
and the kernel is defined on
$H^1(\R^3,\C)$ $\times$ $H^1(\R^3,\C).$

 We show, first, that the kernel
$K(e_{z_1},e_{z_2})$  is effectively
square integrable
(in definition of
\cite[Definition, p.44]{PanPSZ91}),
i.e.,
square integrable
over the sequence of
 finite-\-dimensional subspaces
$L_N,$ .
uniformly  on dimensions of these subspaces.

It is clear that for the kernel
 $K(e_{z_1},e_{z_2})$
 on $ L_N\times
  L_N$
(or on the complex linear span of
$P\Sc_{Re}^{even}({\cal O})$ $\times$
$P\Sc_{Re}^{even}({\cal O})$)
we obtain (with the help
of energy conservation and
 Eq. (4.2) )
 the estimate
$$ |K(e_{z_1},e_{z_2})|
 \leq c e^{{1\over 2}\Vert
T_1 z_1 \Vert^2+{1\over 2} \Vert T_2
z_2\Vert^2}
 \left(\Vert\mu^{1/2}T_1
z_1\Vert+ \Vert\mu^{1/2}T_2
z_2\Vert\right)
 \Vert\mu^{-1/2}h\Vert.
\eqno(4.4)
$$
Here  $\Vert \cdot\Vert$ is the norm of the space
$H^{1/2}(\R^3,\C),$
$$T_j= e^{-s_j\mu}\Lambda(\Sigma,\delta).$$
We remark that the operator
$\Lambda(\Sigma,\delta)$ is a complex linear
operator.

Now we consider integrals of these Wick kernels.
We obtain
a uniform
estimate for squire integrability
of considered Wick kernels over
fi\-ni\-te-\-di\-men\-si\-on\-al
subspaces  $L_N$ of
$H^{1/2}(\R^3,\C)$ and over the corresponding
Gaussian promeasure.
The constructed basis has the standard and
physically natural realification given by the map
$P:\Sc_{Re}({\cal O})\to H^{1/2}(\R^3,\C)$ on
the subspace $\Sc_{Re}^{even}({\cal O}).$
The standard pure real and pure imaginary
subspaces  $L_N$ in chosen  coordinates coincide
with the space  $\R^N$ (and
with the space $i\R^N,$ correspondingly)
as with the Euclidean space.

\medskip
{\bf Remark.}
For  convenience, we chose ${\cal O}$  such, that
${\cal O}^\Theta={\cal O}.$  This means that
 ${\cal O}$ is symmetrical with respect to
the time reflection.
\medskip

Now we consider integrals over finite-\-dimensional
subspaces  $L_N.$
We construct subspaces  $L_N$
with the help of vectors from
$P\Sc_{Re}({\cal O}).$
It follows from
(4.4)
that
$$
\int_{L_{N_1}}\int_{L_{N_2}}\vert
K(e_{z_1},e_{z_2})\vert^2
d\nu(z_1)d\nu(z_2)
$$
\be
&&\leq c\Vert\mu^{-1/2} h\Vert^2
\\&&\times \int_{L_{N_1}}\int_{L_{N_2}}
e^{\Vert T_1 z_1\Vert^2+\Vert T_2 z_2\Vert^2}
(\Vert\mu^{1/2} T_1 z_1\Vert^2
+\Vert\mu^{1/2}T_2 z_2\Vert^2)
d\nu(z_1)d\nu(z_2)
\\&&
\leq c\Vert\mu^{-1/2} h\Vert ^2
\\&&
\times  \Biggl[\left(\int_{L_{N_1}}
e^{p_1 \Vert T_1
z_1\Vert^2}d\nu(z_1)\right)^{1/p_1}
\\&&
\times \left(\int_{L_{N_1}}\Vert\mu^{1/2}
T_1 z_1\Vert^{{2p_1\over p_1-1}}d\nu(z_1)
\right)^{{p_1-1\over p_1}}\int_{L_{N_2}}
e^{\Vert T_2 z_2\Vert^2}d\nu(z_2)
\\&&
+\int_{L_{N_1}}
e^{\Vert T_1 z_1\Vert^2}d\nu(z_1)
\ee
$$\times\left(\int_{L_{N_2}}
e^{p_2\Vert T_2 z_2\Vert^2}
d\nu(z_2)\right)^{1/p_2}
\left(\int_{L_{N_2}}\Vert\mu^{1/2}
T_2 z_2\Vert^{{2p_2\over p_2-1}}
d\nu(z_2)
\right)^{{p_2-1\over p_2}}
\Biggr].
\eqno(4.5)
$$
Here $\Vert\cdot\Vert$
is the norm in the space
$H^{1/2}(\R^3,\C),$ we
use the H{\"o}lder inequality with
$(p_1,q_1)$ for the integral over $d\nu(z_1)$
and for the first item and
with $(p_2,q_2)$ for the integral over $d\nu(z_2)$
and for the second item,
$q_1=p_1(p_1-1)^{-1},$
$q_2=p_2(p_2-1)^{-1},$
and the choice of numbers $p_1,p_2$  would be made later.

Now we consider estimates of integrals in
(4.5).
Consider, first, the integrals
$$
\int_{L_{N_j}}
e^{p_j\Vert T_j z_j\Vert^2}
d\nu(z_j),\quad j=1,2.
$$
We write these integrals
as
$$
\int_{L_N} e^{p\Vert T z\Vert^2}d\nu(z),
\eqno(4.6)
$$
 without subscript  $j.$
Writing these expression  without subscript
we mean that the corresponding letters
have the subscript
$j,$ taking the values 1, or 2.

First of all, on $L_N$
$$
\Vert T z\Vert^2
=\Vert e^{-s\mu}(\mbox{ Re }z+i\mbox{ Im }z)
\Vert^2_{H^{1/2}(\R^3,\C)}
$$
and
$$
\Vert T z\Vert\leq e^{-sm}\Vert z\Vert.
$$
Further,
\be
\Vert T z\Vert^2
&=&\Vert e^{-s\mu}(\mbox{ Re }z +
i\mbox{ Im }z)\Vert^2_{H^{1/2}(\R^3,\C)}\cr
&=&\Vert e^{-s\mu}
\mbox{ Re }z\Vert^2_{H^{1/2}(\R^3,\C)}
+
\Vert e^{-s\mu}
\mbox{ Im }z\Vert^2_{H^{1/2}(\R^3,\C)}\cr
&=&\Vert e^{-s\mu}
\mbox{ Re }z\Vert^2_{H^{1/2}(\R^3)}
+
\Vert e^{-s\mu}
\mbox{ Im }z\Vert^2_{H^{1/2}(\R^3)}.
\ee

The appeared
integrals can be
easily rewritten with the help of introduced variables.
First, the variables over which
we integrate are  the coordinates in the basis
$\{e^{Re}_j, j=1,...,N\}$
(or in the basis
$\{e^{Re}_{j'}, ie^{Re}_{j'}, j'=1,...,N\}$ with the
standard realification), i.e., we have
$z_j = x_j+iy_j
= \langle e^{Re}_j, z\rangle,$ here
$x_j
= \langle e^{Re}_j, \mbox{ Re }z\rangle,$
 $y_j = \langle e^{Re}_j, \mbox{ Im }z\rangle,$
$$ z
 = \sum_j\langle e^{Re}_j, z\rangle e^{Re}_j
= \sum z_j e^{Re}_j
= \sum (x_j e^{Re}_j + y_j ie^{Re}_j),
$$
$$ \mbox{ Re }z
= \sum_j\langle e^{Re}_j,\mbox{ Re } z\rangle
e^{Re}_j
= \sum x_j e^{Re}_j,
$$
$$ \mbox{ Im }z=
\sum_j\langle e^{Re}_j, \mbox{ Im } z\rangle e^{Re}_j
= \sum y_j e^{Re}_j.
$$

For our choice of the basis and the realification
the choice of complex variables in the form
$$
z_j = x_j + iy_j
= \langle e_j, z\rangle
$$
is in agreement with the choice of real
variables in the form
$$
\mbox{ Re }z=
\sum x_j e^{Re}_j =
\sum_j\langle
 e^{Re}_j, \mbox{ Re } z
\rangle e^{Re}_j,
$$
$$
\mbox{ Im }z=
\sum y_j e^{Re}_j =
\sum_j\langle e^{Re}_j,\mbox{ Im } z\rangle e^{Re}_j.
$$
It is convenient to  consider estimates with
the help of these variables because
 contributions of pure real and pure imaginary
variables factorize. In addition,
 for our choice  of the basis and the localization operator
 (see Section 5)
$$ z
= \sum (x_j e^{Re}_j + iy_j e^{Re}_j),
= \Lambda(\Sigma,\delta)z
$$ $$
= \sum (x_j
\mu^{-1}\Lambda\mu e^{Re}_j +
 iy_j \mu^{-1}\Lambda\mu e^{Re}_j)
 $$
 and
$$
\Vert z\Vert^2
=\sum^N_{j=1}(x^2_j+y^2_j),
$$
$$
 \Vert Tz\Vert^2
=\Vert e^{-s\mu}(\mbox{ Re }z +
i\mbox{ Im }z)\Vert^2_{H^{1/2}(\R^3,\C)}
$$
$$
=\Vert e^{-s\mu}
\mbox{ Re }z\Vert^2_{H^{1/2}(\R^3)}
+\Vert e^{-s\mu}\mbox{ Im }z
\Vert^2_{H^{1/2}(\R^3)}
$$
$$
=\Vert e^{-s\mu}\sum x_j
e^{Re}_j\Vert^2_{H^{1/2}(\R^3)}
+\Vert e^{-s\mu}\sum y_j
e^{Re}_j  \Vert^2_{H^{1/2}(\R^3)}
$$ $$
=\Vert
e^{-s\mu}\mu^{-1}\Lambda\mu\sum x_j
e^{Re}_j
\Vert^2_{H^{1/2}(\R^3)}
 + \Vert
e^{-s\mu}\mu^{-1}\Lambda\mu\sum y_j e^{Re}_j
\Vert^2_{H^{1/2}(\R^3)}
$$
$$
=\sum^{N}_{j,j'=1}x_j x_{j'}K^{Re}_{jj'}
 + \sum^{N}_{j,j'=1}
y_j y_{j'}K^{Im}_{jj'},
\eqno(4.7)
$$
where
$$
K^{Re}_{jj'}= K^{Im}_{jj'}
=\langle e^{-s\mu}\mu^{-1}\Lambda\mu e^{Re}_j,
e^{-s\mu}
\mu^{-1}\Lambda\mu
e^{Re}_j\rangle_{H^{1/2}(\R^3)}
$$ $$=(e^{Re}_j,\mu\Lambda\mu^{-1}
e^{-2s\mu}\Lambda\mu
e^{Re}_{j'})_{L_2(\R^3)}.
$$
 We emphasize once more that these equalities are
 fulfilled on the subspace
 $L_N$ only. A sequence of these subspaces is dense in
$H^1(\R^3)$ and the equalities are written with the help
 of localization operators.

The same expression can be rewritten as
matrix elements of fi\-ni\-te-\-di\-men\-sion\-al operator
$$
K^{Re}_N =
P^{Re}_N\mu\Lambda\mu^{-1}
e^{-2s\mu}\Lambda\mu P^{Re}_N,
$$
$$
K^{Im}_N =
P^{Im}_N\mu\Lambda\mu^{-1}
e^{-2s\mu}\Lambda\mu P^{Im}_N,
$$
Here $P^{Re}_N$  is the projection on
the subspace  $L^{Re}_N=\mbox{ Re }L_N,$
which is a finite--dimensional real subspace of
$P\Sc^{even}_{Re}({\cal O})$
and  $P^{Im}_N$  is the projection on
the subspace  $L^{Im}_N=i\mbox{ Im }L_N,$
which is a finite-\-dimensional real linear subspace of
$P\Sc^{odd}_{Re}(\R^4),$
it can be written also as a subspace of
$\mu^{-1}P{\partial\over\partial t} \Sc^{even}_{Re}(\cal O).$

The following inequality is fulfilled on
$L_N$
$$
\Vert Tz\Vert\leq e^{-sm}\Vert z\Vert.
$$
This inequality, or direct consideration
of real subspaces, allows to obtain the
inequalities
\be
\sum^N_{j,j'=1}x_j x_{j'}K^{Re}_{jj'}
&=&
\Vert
e^{-s\mu}\mu^{-1}\Lambda\mu
\sum^N_{j=1}x_j e^{Re}_j
\Vert^2_{H^{1/2}(\R^3)}
\cr
&=&\Vert
e^{-s\mu}\mu^{-1}\Lambda\mu \mbox{ Re }z
\Vert^2_{H^{1/2}(\R^3)}
\cr
&=&\Vert
e^{-s\mu} \mbox{ Re }z
\Vert^2_{H^{1/2}(\R^3)}
\cr
&\leq&
e^{-2sm}\Vert\mbox{ Re }z\Vert^2_{H^{1/2}(\R^3)}
= e^{-2sm}\sum^N_{j=1}x^2_j,
\ee
\be
\sum^N_{j,j'=1}y_j y_{j'}K^{Im}_{jj'}
&=&
\sum^N_{j,j'=1}y_j y_{j'}K^{Re}_{jj'}
\cr &\leq&
 e^{-2sm}\sum^N_{j=1}y^2_j.
\ee
This implies that
$K^{Re}_{jj'}$ is  symmetric
positive matrix, thus, it is
diagonalizable (by orthogonal transformation)
and has positive eigenvalues.
The estimates imply that the
eigenvalues of the matrix
 $\{K^{Re}_{jj'}\}$
are not greater than $e^{-2sm}.$
Moreover, the explicit form of the matrix
and the explicit form of the operator
that gives this matrix imply  that
this operator is an
 operator of trace class and
 the sum of eigenvalues of the matrix
$\{ K^{Re}_{jj'}\}$
is uniformly bounded on dimension of
subspaces $L_N.$

\medskip
{\bf Remark.}
 We remark that the considered operator  $T$
(due to the use of localization operator
and the consideration of real
subspace and pure  imaginary subspace of
  the complex linear Hilbert space)
has the form
$$
e^{-s\mu}R\Lambda_{\C}(\Sigma({\cal O}),\delta)R^{-1},
 $$
 or
$$
e^{-(s-|y|)\mu}R\Lambda_{\C}(\Sigma({\cal O}),\delta)R^{-1}
 $$
 for the general case. The operator $T$ acts in the complex
Hilbert space $H^1(\R^3,\C)$ and it is a complex
linear operator.

\medskip
\medskip
We continue the estimate
of the Wick kernel  $K(e_{z_1},e_{z_2})$
and integrals appeared
in
(4.5).

We note that the operators
$K^{Re}_{N},$ $K^{Im}_{N}$ are defined by the
fi\-ni\-te--\-di\-men\-si\-on\-al
projection of the operator $T^*T$
on the finite dimensional subspace
$$\mbox{Re}L_N\oplus i\mbox{Im} L_N,$$
with our choice of the basis
this projection is given by the matrix
$$\pmatrix{K^{Re}_{jj'}&0\cr 0&K^{Re}_{jj'}\cr}.$$

Now taking into account that the eigenvalues
of the matrix
$K^{Re}_{jj'},$
 are less than
$e^{-2sm},$ or that
the norm of the operator $T,$ restricted on
the subspaces $L_N,$  is less than
$e^{-sm},$ $\Vert T|_{L_N}\Vert\leq e^{-sm},$
 we
choose
$p = (1 + (1-e^{-sm/2}))^2
= (2-e^{-sm/2})^2$
for the integral  (4.6).
Then
$ p^{1/2}\Vert T\Vert
\leq e^{-sm/2}$
(because $2\leq e^x+e^{-x}$
 and  $p^{1/2}=2-e^{-sm/2}\leq e^{sm/2}$).

Using this value of $p$
   and making the change of
variables, we obtain
for the integral
$$
\int_{L_N}e^{p\Vert T z\Vert^2} d\nu(z),
$$
 the following equality
$$
\int_{L_N}e^{p\Vert T z\Vert^2} d\nu(z)
=
\det(1-pK^{Re}_N)^{-1/2}
\det(1-pK^{Im}_N)^{-1/2}
$$ $$= \prod^N_{j=1}(1-p\lambda^{Re}_{j,N})^{-1}.
$$
Here
$\lambda^{Re}_{j,N}$
 are eigenvalues of the operator
$$
K^{Re}_N
= P^{Re}_N\mu\Lambda\mu^{-1}
e^{-2s\mu}\Lambda\mu P^{Re}_N.
$$
The eigenvalues of the operator $K^{Im}_N$
are the same and are equal to $\lambda^{Re}_{j,N}.$
The eigenvalues are
defined by the real symmetric positive matrix
$$\{K^{Re}_{jj'}, j,j'=1,...,N\}.$$

The eigenvalues
$\lambda^{Re}_{j,N}$
are positive and the trace
 of the operators
 $K^{Re}_N$ and  $K^{Im}_N$
is bounded uniformly on dimension  $N$
of the space $L_N,$
$$
\sum^N_{j=1} \lambda^{Re}_{j,N}
= \mbox{Tr } K^{Re}_N
\leq
\mbox{Tr }(\mu\Lambda\mu^{-1}
e^{-2s\mu}\Lambda\mu)
$$
$$
=\Vert e^{-s\mu}
\mu^{-1}\Lambda\mu\Vert^2_{HS(H^{1/2}(\R^3))}
 = \Vert
e^{-s\mu}\mu^{-1/2}
\Lambda\mu\Vert^2_{HS(L_2(\R^3))}, \eqno(4.8)
$$
 Therefore, $$
\int_{L_N}
e^{p\Vert T z\Vert^2} d\nu(z)=
\prod^N_{j=1}(1-p\lambda^{Re}_{j,N})^{-1}
$$ $$
\leq
\exp(-\sum^N_{j=1}\ln(1-p\lambda^{Re}_{j,N}))
$$
$$
\leq\exp(\sum^N_{j=1}
{p\lambda^{Re}_{j,N}\over 1-p\max_j\lambda^{Re}_{j,N}}).
$$
We have used the inequality
$$
-\ln(1-a)\leq{1\over 1-a_0}a\;\;
\mbox{ for }\;\;0\leq a\leq a_0<1,
$$
(this inequality is implied by the equality
for $a=0$ and by the positive derivative).

We note, further, that
 $$
p\max_j\lambda^{Re}_{j,N} =
p\Vert K^{Re}_N\Vert
\leq p\Vert T\Vert^2
\leq e^{-sm}$$
and
 \be (1-p\max_j\lambda_{j,N})^{-1}
&=&
(1-p^{1/2}\max_j\lambda^{1/2}_{j,N})^{-1}
(1+p^{1/2}\max_j\lambda_{j,N}^{1/2})^{-1}
\cr
&\leq&
(1-e^{-sm/2})^{-1}\leq 2+{4\over ms},
\ee
because $(1-e^{-x})^{-1} \leq 2 + 2x^{-1}.$
The last inequality is implied by the
inequality
$$1 - e^{-x} \geq {1\over2}\geq (2+{2\over x})^{-1},$$
the first part of which is equivalent to the inequality
 $x\geq \ln 2,$ and by inequalities
 $$1 - e^{-x} \geq {1\over 2}x
\geq (2+{2\over x})^{-1}$$ for $0\leq x\leq \ln 2.$
The last inequalities are implied by relation
$$
1-e^{-x}=-\int^1_0{d\over ds}
e^{-sx}dx=\int^1_0 e^{-sx}xds
$$
$$
\geq x\int^1_0 e^{-s\ln 2}ds=
x(2\ln 2)^{-1}\geq{x\over 2}\geq(2+{2\over x})^{-1}.
$$

In addition, inequality  (4.8) and inequality (4.9)
imply, that
$$
\sum_j\lambda^{Re}_{j,N} \leq
\Vert
e^{-s\mu} \mu^{-1}\Lambda \mu
\Vert^2_{HS(H^{1/2}(\R^3))} $$
$$
\leq c_1 s^{-2\alpha}
\Vert\mu^{-\alpha-1}
\Lambda\mu\Vert^2_{HS(H^{1/2}(\R^3))}
\leq c_2 s^{-2\alpha}
$$ for
$\Vert\mu^{-\alpha-1}\Lambda\mu
\Vert_{HS(H^{1/2}(\R^3))} < \infty.$
For sufficiently large $\alpha$ the
boundedness of Hil\-bert-\-Schmidt norm
 of the operator
 $$
\mu^{-\alpha-1}\Lambda \mu$$
 is evident and conditions on the
parameter $\alpha$ are considered in Appendix.

Thus,
for $\Vert\mu^{-\alpha-1}\Lambda
\mu\Vert_{HS(H^{1/2}(\R^3))}< \infty$
(for instance, for $\alpha>2,$ see Appendix)
$$
\sum^N_{j=1}{p\lambda^{Re}_{j,N}
\over 1-p\max_j\lambda^{Re}_{j,N}}
\leq
4(2+{4\over ms})\sum^N_{j=1}\lambda^{Re}_{j,N}
\leq c_3(1+s^{-2\alpha-1})
$$
and, so
$$
\int_{L_N} e^{p\Vert Tz\Vert^2}d\nu(z)\leq
e^{c(1+ s^{-2\alpha-1})}
$$
for
$\Vert\mu^{-\alpha-1}
\Lambda\mu \Vert_{HS(H^{1/2}(\R^3))}< \infty.$

Further, we consider the integral
$$
(\int_{L_N}\Vert \mu^{1/2}Tz
\Vert^{{2p\over p-1}}
d\nu(z))^{{p-1\over p}}\eqno(4.9)
$$
with our choice of the parameter
$p=(2-e^{-sm/2})^2.$
We use the bounds
$$\Vert \mu^{1/2}Tz\Vert \le s^{-1/2}
\Vert e^{s\mu/2} Tz \Vert
$$
 and $$
\Bigl(\int_{L_N}\Vert Bz\Vert^{2n} d\nu(z)\Bigr)^{1/2n}
\le  \Bigl({(2n)!\over\beta^{2n}} \int_{L_N}
 \exp(\beta \Vert Bz\Vert^2) d\nu(z)\Bigr)^{1/2n}
 $$ $$
\le c{n\over \beta}
 \Bigl(\int_{L_N}
 \exp(\beta \Vert Bz\Vert^2) d\nu(z)\Bigr)^{1/2n},
 $$
 here $B$ is a linear operator.

   First, we obtain that
$$
(\int_{L_N}\Vert \mu^{1/2}Tz
\Vert^{{2p\over p-1}}
d\nu(z))^{{p-1\over p}}
\le s^{-1}
 ( \int_{L_N}
 \Vert e^{s\mu/2} Tz\Vert^{2n_0} d\nu(z))^{1/n_0}.
$$
 Here $e^{s\mu/2}T$ is correctly defined
 and $n_0$ is chosen as the smallest integer greater than
 $p(p-1)^{-1}.$
Since $p=(2-e^{-sm/2})^2$ it is clear
that $p(p-1)^{-1}
 \le n_0\le 4+4m^{-1}s^{-1}.$

The estimate for (4.9) has the following form
\be
(4.9)&\le&
 s^{-1}( \int_{L_N}
 \Vert e^{s\mu/2} Tz\Vert^{2n_0} d\nu(z))^{1/n_0}
\\ &\le&
  c \,s^{-1}{n^2_0\over \beta^2}
 \Bigl( \int_{L_N}
 \exp(\Vert e^{s\mu/2} Tz\Vert^2) d\nu(z)\Bigr)^{1/n_0}.
\ee
$$ \le  c_1 (s^{-1}+m^{-1}s^{-2})
  \beta^{-2}
 \Bigl( \int_{L_N}
 \exp(\Vert e^{s\mu/2} Tz\Vert^2) d\nu(z)\Bigr)^{1/n_0}.
\eqno(4.10)$$
 The appeared integral is expressed with
the help of the operator
 $$ K^{Re}_{1,N}=P^{Re}_N
\mu\Lambda\mu^{-1}e^{-s\mu}\Lambda\mu
 P^{Re}_N, $$
 and is equal to
$$
\int_{L_N}\exp(\beta\Vert e^{s\mu/2} T z\Vert^2) d\nu(z)
=
\det(1-\beta K^{Re}_{1,N})^{-1}
$$
 Taking, for instance, $\beta=e^{-m}$
 (here $m$ is the mass constant) we obtain the estimate
 $$ \mbox{R.h.s.} (4.10) \le c_2(1+s^{-2})
 \exp(c_3n_0^{-1}(1+s^{-2\alpha -1}))
 \le c_4\exp(c_4(1+s^{-2\alpha -1})).
 $$

 Now we use (4.5) and
 apply the derived estimates.
 We obtain the following uniform estimate
 $$ \sup_{L_{N_1},L_{N_2}}
\bigl( \int_{L_{N_1}}
 \int_{L_{N_2}}
 \vert  K(z_1,z_2)\vert^2
d\nu(z_1)d\nu(z_2)\bigr)^{1/2}
$$
$$
\leq c_1\exp(c s_1^{-2\alpha-1}+c s_2^{-2\alpha-1})
 \Vert\mu^{-1/2}h\Vert,
$$
$\alpha$ is such that
$\Vert\mu^{-\alpha-1}\Lambda\mu
\Vert_{HS(H^{1/2}(\R^3))}<\infty,$
for instance, $\alpha>2$ (see Appendix).
This estimate, Lemma 1.1 and Lemma 1.5
\cite[Lemma 1.1, Lemma 1.5]{PanPSZ91}
imply that $K(e_{z_1},e_{z_2})$ is the
Wick kernel of a unique Hilbert-\-Schmidt operator
(in the Fock space). In addition, it follows that
 $$ \left\vert
 \int_{L_{N_1}}
 \int_{L_{N_2}}
 \overline{\chi_1({\bar z}_1)}
 K(z_1,z_2)
\chi_2({\bar z}_2)
d\nu(z_1)d\nu(z_2)\right\vert
$$
 $$ \le \Bigl( \int_{L_{N_1}}
 \vert \chi_1({\bar z}_1) \vert^2 d\nu(z_1)\Bigr)^{1/2}
  \Bigl( \int_{L_{N_2}}
 \vert \chi_2({\bar z}_2) \vert^2 d\nu(z_2)\Bigr)^{1/2}
$$ $$ \left\vert
  \int_{L_{N_1}}
  \int_{L_{N_2}}
 \vert  K(z_1,z_2)\vert^2
d\nu(z_1)d\nu(z_2)\right\vert^{1/2}
$$
$$
\leq c_1\exp(c s_1^{-2\alpha-1}+c s_2^{-2\alpha-1})
\;\Vert\chi_1\Vert_{\Phi o\kappa}\;
\Vert\chi_2\Vert_{\Phi o\kappa}\;
 \Vert\mu^{-1/2}h\Vert,
$$
$\alpha$ is such that
$\Vert\mu^{-\alpha-1}
\Lambda\mu\Vert_{HS(H^{1/2}(\R^3))}<\infty,$
for instance, $\alpha>2$ (see Appendix),
and $\Vert . \Vert_{\Phi o\kappa}$ is the norm in the
Fock space
${\Phi o\kappa}.$
The obtained estimate implies also that
 $$ \sup_{L_{N_1},L_{N_2}}
\bigl( \int_{L_{N_1}}
 \int_{L_{N_2}}
 \vert  \phi(t_1+is_1,t_2+is_2,e_{z_1},e_{z_2})\vert^2
d\nu(z_1)d\nu(z_2)\bigr)^{1/2}
$$
$$
\leq c_0
 \Vert\mu^{-1/2}h\Vert
\exp(c s_1^{-2\alpha-1}+c s_2^{-2\alpha-1}).
$$
This estimate, Lemma 1.1 and Lemma 1.5
\cite[Lemma 1.1, Lemma 1.5]{PanPSZ91}
imply the existence of a
unique Hilbert-\-Schmidt operator
in the Fock space and with the Wick kernel equal
 to the Wick kernel
 $\phi(t_1+is_1,t_2+is_2,e_{z_1},e_{z_2}).$

Therefore, Theorem 4.2 is proved for the case
 of the interacting  field.
 The $out$-going field $\phi_{out}$
 can be considered analogously and analogously
 we obtain the same estimates.
 This completes the proof of Theorem 4.2.
\medskip

To prove Theorem 4.3 we consider the general case
 of the kernels
 $$K(e_{z_1},e_{z_2})$$ with
 $(t_1+is_1,x_1+iy_1),$
 $(t_2+is_2,x_2+iy_2)$ $\in$ $\R^4+iV_+.$
 To consider the
 general case
 of the kernels
with $(t+is,x+y)$ $\in$ $\R^4+iV_+$ we
 use the bound
$$ s(-\Delta+m^2)^{1/2}\pm y i \mbox{ grad }
\ge (s-|y|)(-\Delta +m^2)^{1/2},
$$ i.e.,
$$
 s\mu(k) -y k \le (s-|y|) \mu(k),
$$
and the estimate (4.4) for the general case written
with the help of operators $T_1,$ and $T_2$ and
Equality (4.3).

This allows to obtain analogously the estimate
 $$ \sup_{L_{N_1},L_{N_2}}
\bigl( \int_{L_{N_1}}
 \int_{L_{N_2}}
 \vert  K(z_1,z_2)\vert^2
d\nu(z_1)d\nu(z_2)\bigr)^{1/2}
$$
$$
\leq c_1\exp(c (s_1-|y_1|)^{-2\alpha-1}
+ c (s_2-|y_2|)^{-2\alpha-1})
 \Vert\mu^{-1/2}h\Vert,
$$
where $\alpha$ is such that
$\Vert\mu^{-\alpha-1}\Lambda\mu
\Vert_{HS(H^{1/2}(\R^3))}<\infty,$
for instance, $\alpha>2.$

This estimate
 implies the estimate
 $$ \sup_{L_{N_1},L_{N_2}}
\bigl( \int_{L_{N_1}}
 \int_{L_{N_2}}
 \vert
\phi(t_1+is_1,x_1+iy_1,t_2+is_2,x_2+iy_2,
 e_{z_1},e_{z_2})\vert^2
d\nu(z_1)d\nu(z_2)\bigr)^{1/2} $$
 $$ \leq c_1
 \Vert\mu^{-1/2}h\Vert
\exp(c s_1^{-2\alpha-1}+c s_2^{-2\alpha-1})
 \|\mu^{-1/2}h\|
$$
and  Lemma 1.1, Lemma 1.5
\cite[Lemma 1.1, Lemma 1.5]{PanPSZ91}
imply the existence of a
 unique Hilbert-\-Schmidt operator
(in the Fock space of the $in$-filed) with the
 Wick kernel coinciding with
$$\phi(t_1+is_1,x_1+iy_1,t_2+is_2,x_2+iy_2,
 e_{z_1},e_{z_2}).$$
 The analogous assertion is also valid for the
 $out$-field $\phi_{out}.$
 Theorem 4.3 is proved.
\medskip

 The proved assertion about square integrability of Wick
 kernels means that the expressions
  $$ \langle\chi_1,e^{-s_1 H}
 \phi_\#(t,h)e^{-s_2 H}\chi_2\rangle$$
 is correctly defined and satisfy the estimate
  $$\vert \langle\chi_1,e^{-s_1 H}
 \phi_\#(t,h)e^{-s_2 H}\chi_2\rangle\vert$$
 $$ \leq c_1\exp(c
 s_1^{-2\alpha-1} + c s_2^{-2\alpha-1})
\;\Vert\chi_1\Vert_{\Phi o\kappa}\;
 \Vert\chi_2\Vert_{\Phi o\kappa}\;
 \Vert\mu^{-1/2}h\Vert,
$$
where $\alpha$ is such, that
$\Vert\mu^{-\alpha-1}\Lambda\mu
 \Vert_{HS(H^{1/2}(\R^3))}<\infty,$
 for instance, $\alpha>2.$

\medskip
We formulate these assertions as the separate theorems.

\medskip
{\bf Theorem 4.4.}
 {\it
For the operator
$e^{-s_1 H}\phi(t,h)e^{-s_2 H},$ $s_1>0,$
$ s_2>0,$ the following estimate is valid
$$
\vert\langle\chi_1,e^{-s_1 H}
\phi(t,h)e^{-s_2 H}\chi_2\rangle\vert
$$
$$
\leq c_1\exp(c s_1^{-2\alpha-1}+c s_2^{-2\alpha-1})
\;\Vert\chi_1\Vert_{\Phi o\kappa}\;
\Vert\chi_2\Vert_{\Phi o\kappa}\;
\; \Vert\mu^{-1/2}h\Vert,
$$
where $\alpha$ is such, that
$\Vert\mu^{-\alpha-1}
\Lambda(\Sigma,\delta)\mu
\Vert_{HS(H^{1/2}(\R^3))}<\infty,$
for instance, $\alpha > 2.$
Here $\chi_1,$ $\chi_2$
are vectors in the Fock space
$\Phi o\kappa$
with
Fock space norm, for
$h$ we  use the norm of the space
 $H^{1/2}(\R^3),$ and
$\Vert\cdot\Vert_{HS(H^{1/2}(\R^3))}$
is the Hil\-bert-\-Schmidt norm for  operators
in the Hilbert space $H^{1/2}(\R^3).$

 The analogous assertion is valid for the quantum field
 $\phi_{out}.$
}

\medskip
{\bf Theorem 4.5.}
 {\it
Let
$s_1>|y_1|, $
$s_2>|y_2|,$ i.e. $(s_1, y_1),$ $(s_2, y_2)$
$\in$ $V_+.$ For the operator
$  e^{-s_1 H-y_1 P}
\phi(t,h)e^{-s_2 H-y_2 P}$
the following
estimate is valid
$$ \vert\langle\chi_1, e^{-s_1 H-y_1 P}
\phi(t,h)e^{-s_2 H-y_2 P}\chi_2\rangle\vert
$$
$$
\leq c_1\exp(c (s_1-|y_1|)^{-2\alpha-1}
+ c (s_2-|y_2|)^{-2\alpha-1})
\,\Vert\chi_1\Vert_{\Phi o\kappa}\;
\Vert\chi_2\Vert_{\Phi o\kappa}\;
 \Vert\mu^{-1/2}h
\Vert,
$$
where $\alpha$ is such, that
$\Vert\mu^{-\alpha-1}
\Lambda\mu
\Vert_{HS(H^{1/2}(\R^3))}<\infty,$
for instance, $\alpha > 2$ (see Appendix).
Here $\chi_1,\chi_2$
are vectors in the Fock space
$\Phi o\kappa$
with
Fock space norm, for
$h$ we use the norm of the space
 $H^{1/2}(\R^3),$ and
$\Vert\cdot\Vert_{HS(H^{1/2}(\R^3))}$
is the Hil\-bert-\-Schmidt norm for operators
in the Hilbert space  $H^{1/2}(\R^3).$

The analogous assertion is valid for the quantum field
 $\phi_{out}.$
}



\section*{5. Operator of localization.}

Now  we consider special subspaces of
localized vectors in the complex Hilbert space
$H^{1/2}(\R^3,\C).$
We are required these vectors
in order to obtain uniform estimates on
Wick kernels and for the consideration of
coherent vectors of the form
$$
\exp(\phi_{in}^+v)\Omega
=\exp(\sum_k a^*_k \langle
\varphi_k, Pv\rangle_{H^{1/2}(\R^3,\C)})\Omega,
$$ where
$$
(\phi^+_{in}v)=\int\phi^+_{in}(t,x)v(t,x)dt dx
$$
$$
=\int\phi^+_{in}(0,x)\int e^{i\mu t}v(t,x) dt dx=
 \sum_k a^*_k\langle \varphi_k,Pv\rangle_{H^{1/2}(\R^3,\C)}
$$
and $Pv \in H^{1/2}(\R^3,\C).$
Here we use the representation of the free
$in$-field in terms of annihilation $a_k$
and creation $a^*_k$ operators and the
orthonormal basis $\varphi_k$ in
$H^{1/2}(\R^3,\C),$ see
\cite[Sect. 3]{Osi96b}.

We remark  that the function
$\exp(\langle z,P
v\rangle_{H^{1/2}(\R^3,\C)})$  (a coherent kernel
in the representation of the Fock space of the
$in$-field with the help of entire
antiholomorphic functions)
corresponds to the vector
$\exp(\phi_{in}^+v) \Omega,$
or to the vector
$1\oplus^\infty_{n=1}
n!^{-1/2} Pv \otimes...\otimes Pv $
for the standard representation of the Fock space.

If  $z\in H^{1/2}(\R^3,\C),$ then  a unique free solution
$$
\cos\mu t\mbox{ Re }z+\sin\mu t\mbox{ Im }z
\quad (= \mbox{ Re }(e^{-i\mu t}z))
$$
corresponds  to this vector.
This free solution has
initial data  (at the moment $t=0$) equal to
$(\mbox { Re }z,\; \mu\mbox{ Im }z)$
and the positive frequency part of solution
$z^+ (t) = \overline{e^{-i\mu t}z},$
$z^+(0)=\overline{z}.$

We define the map $P$ of the Schwartz space
 $\Sc_{Re}(\R^4)$ of real-\-valued test functions
into the complex Hilbert space
$H^{1/2}(\R^3,\C).$
This map is generated by positive
frequency part of free solutions.
In Goodman's notation \cite{Goo64}
this map is the projection operator of
the space $\Sc_{Re}(\R^4)$ into $H^{1/2}(\R^3,\C).$

\medskip
{\bf Remark.}
  For the Wightman reconstruction theorem, see
\cite{Jos65},
\cite{BogLOT87},
 this map corresponds
to the transition from the topological
algebra  $\underline\Sc$ to vectors
 $\prod\phi(f_j)\Omega$
in the physical Hilbert space.
\medskip

By definition, if
 $f \in \Sc_{Re}(\R^4),$ then for any
$z\in H^{1/2}(\R^3,\C)$
\be
\langle z,Pf\rangle_{H^{1/2}(\R^3,\C)}
&=& \int z^+(t,x) f(t,x)dtdx
\cr
&=&
\int\overline{e^{-i\mu t}z(x)}f(t,x)dtdx
\cr
&=&
\int\overline{z(x)}\int e^{i\mu t}f(t,x)dtdx
\cr
&=&\langle z,\int\mu^{-1}
e^{i\mu t}f(t,\cdot)dt\rangle_{H^{1/2}(\R^3,\C)}.
\ee
It is clear that
\be
(Pf)(x)
&=&\int(\mu^{-1}e^{i\mu t}f)(t,x)dt
\cr
&=&\int W^0_2(t',x'-x)f(t',x')dt'dx',
\ee
where the generalized function
$$
W^0_2(t,x)=\int e^{i\mu t}e^{-ipx}{d^3 p\over 2\mu}
=\int\vartheta(p_0)\delta(p^2-m^2)e^{ip_0 t-ipx}d^4 p
$$
coincides with the two--point Wightman function
of the free quantum field.

We note, that the image of the map $P$
belongs to  $\Sc(\R^3,\C),$
$$P\Sc_{Re}(\R^4)\subset\Sc(\R^3,\C).$$

Therefore for  $v\in \Sc_{Re}(\R^4)$
 $\;\; \exp(\phi^+_{in}v)\Omega=e_{Pv}, $
where
\be
(\phi^+_{in}v)
&=&\int\phi^+_{in}(t,x)v(t,x)dtdx\cr
&=&\int\phi^+_{in}(0,x)
\int(e^{i\mu t}v)(t,x)dtdx\cr
&=&\sum_k a^*_k \varphi_k(x)
\int (e^{i\mu t} v)(t,x)dtdx \cr
 &=&\sum_k a^*_k \langle \varphi_k,
 Pv\rangle_{H^{1/2}(\R^3,\C)},
\ee
and the coherent vector $e_{Pv} $
 is the entire antiholomorphic function,
$$  e_{Pv} = \exp (\langle z, Pv\rangle_{H^{1/2}(\R^3,\C)}).
$$
Here we use the representation of the free
$in$-field in terms of annihilation $a_k$
and creation $a^*_k$ operators and the
orthonormal basis $\varphi_k,$  see
\cite[Sect. 3]{Osi96b}.

Now we consider the map
$
P :\Sc_{Re}(\R^4)
\to H^{1/2}(\R^3,\C),$
$$
(Pf)(x)
=\int(\mu^{-1}e^{i\mu t}f)(t,x)dt
= \int W^0_2(t',x'-x)f(t',x')dt'dx'.
$$
First, we note that the image of  $P$
$$
\{Pf\;|\;f\in\Sc_{Re}(\R^4)\}
$$
is dense in the domain
$D(\mu^{1/2})=H^1(\R^3,\C)\subset H^{1/2}(\R^3,\C).$
Really,
$$
(Pf)(x) = \int(\mu^{-1}\cos\mu t f)(t,x)dt
 + i \int(\mu^{-1}\sin\mu t f)(t,x)dt.
$$
Let $f(t,x) = \sigma(t)(\mu f_1)(x) -
{\partial\over\partial t}\sigma(t)
f_2(x),$
then, for instance, for
$f_1, f_2 \in \Sc_{Re}(\R^3)$ and $\sigma(t)$
with compact support, even,
$\sigma(t)= \sigma(-t),$ convergent to the
$\delta$--function
\be
(Pf)
&=&\int\cos\mu t\sigma(t)dt \;f_1
- i\int\mu^{-1}\sin\mu t
{\partial\over\partial t}\sigma(t)dt \;f_2
\\&=&\int\cos\mu t\sigma(t)dt \;f_1
+ i\int\cos\mu t\sigma(t)dt \;f_2
\\&\to& f_1+if_2
\ee
in the sense of convergence in
$\Sc(\R^3,\C)$. But the  set of such
vectors is dense in  $\Sc(\R^3,\C)$ and,
thus, is dense in the space $H^1(\R^3,\C)$ and in
 the space $H^{1/2}(\R^3,\C).$

\medskip
Now we show that the image of the transformation
$P$  restricted on functions with compact support
generates a dense subspace in
$H^{1/2}(\R^3,\C).$

Let ${\cal O}$ be an open set in  $\R^4$  and
$$
\Sc_{Re}({\cal O})=
\{f\in\Sc_{Re}(\R^4)\;\vert\;
\mbox{ supp }f\subset{\cal O}\}.
$$
The following lemma is valid.

\medskip
{\bf Lemma 5.1.}
 {\it
$P\Sc_{Re}({\cal O})$
is dense in $H^n(\R^3,\C),$  in particularly,
$P\Sc_{Re}({\cal O})$
is dense in the space $H^1(\R^3,\C)$ and in the space
 $H^{1/2}(\R^3,\C).$ }

\medskip
{\bf Remarks.}
  1. This is, in fact, an assertion of the type of
the Reeh-\-Schlieder theorem
 \cite{Jos65}
(and it is a consequence of the
theorem of Reeh-\-Schlieder
\cite{Jos65}
applied to the case of the free
quantum field)
or of the type of the assertion of Goodman for
solutions of the free wave equation
\cite[Corollary 3.1]{Goo64}.

2. For the proof of the theorem it was possible
to use the scheme of the proof of the Reeh--Schlieder
theorem, given in Jost's book
\cite{Jos65}.
 The another way consists in the using  the
 fact that  the space
$H^{1/2}(\R^3,\C)$ is the one-\-particle subspace
in the free Fock space,
the  $n$--particle subspaces are orthonormal for
$n \not = n'$ and the vector
$\phi_{in}(f)\Omega$ in the free Fock space corresponds namely
 to the vector
$Pf.$

\medskip
{\it Proof of Lemma  5.1.}
First, we  consider the case
of the complex Hilbert
space
 $H^{1/2}(\R^3,\C).$ It is sufficient
to show, that if
$\psi\in H^{1/2}(\R^3,\C)$ and $\psi$ are orthonormal to
$P\Sc_{Re}({\cal O})$
in $H^{1/2}(\R^3,\C),$ then $\psi=0.$

If  $\psi$ is orthonormal to
$P\Sc_{Re}({\cal O})$ in $H^{1/2}(\R^3,\C),$
then for all
$f\in \Sc_{Re}({\cal O})$ the following equality is fulfilled
$$
\langle\psi, P f\rangle_{H^{1/2}(\R^3,\C)} = 0.$$

Since this equality is continuous in
$f\in\Sc_{Re}(\R^4),$ it implies that the
generalized function
$$
F(\psi)(t,x) = \int W^0_2(t,x'-x)\psi(x')dx',
\quad(t,x)\in\R^4, $$
is equal to zero in ${\cal O}.$

The properties of the free two-\-point function
imply that this generalized function have the
 analytical continuation
$F(\psi)(t+is,x+iy)$ into $(s,y)\in V_+.$

The ``edge of the wedge'' theorem of Bogoliubov
gives an analytical continuation of  $F(\psi)$
through  ${\cal O}$
into $(s,y)\in V_-,$
\cite{BogMP58,StrW64}.
We
refer to the
 ``edge of the wedge'' theorem formulated,
for instance, in  Vladimirov's book
 \cite[\S 27.1, p. 295]{Vla66}.
To apply this theorem we choose appropriate
sets and functions.
As an open set we take the set
$\{(t+is,x+iy) \;\vert\; (s,y)\in V_+\cup V_-\},$
an open cone  $C$ is
 $V_+ \cup V_-,$
in this case  $-C = C.$ As an open set
 in $ \R^4$
 we take  ${\cal O}.$
As a holomorphic function we take
the function
$F(\psi)(t+is,x+iy)$ for
$$\{(t+is,x+iy) \;\vert \;(s,y)
\in V_ + \cup V_-\}
\cap \{(t+is,x+iy) \;\vert \;(s,y)\in V_+\},
$$
and equal to zero for
$$\{(t+is,x+iy) \;\vert \;(s,y)
\in V_ + \cup V_-\}
\cap \{(t+is,x+iy) \;\vert \;(s,y)\in V_-\}.
$$
By
the ``edge of the wedge'' theorem
the function
$F(\psi)(t+is,x+iy)$ is holomorphic
and single-valued in the domain
$\{(t+is,x+iy) \;\vert \;(s,y) \in V_+\cup V_-\}
\cup\tilde{\cal O},$ where
$\tilde{\cal O}$ is a complex neighborhood of the set
${\cal O}.$ Therefore, an analytical continuation
gives zero. But then
$F(\psi) = 0$ and, thus,
$\psi \in \{Pf \;\vert\; f \in \Sc_{Re}(\R^4)\}^\perp
= \{0\},$
the result that we need to prove.
\medskip
Since for
$\psi\in H^n(\R^3,\C)$
$$
\langle\psi, Pf\rangle_{ H^n(\R^3,\C)}
=\langle\psi, \mu^{2n-1}Pf
\rangle_{H^{1/2}(\R^3,\C)}
$$
the space
$H^n(\R^3,\C)$ is considered analogously
and in the same way we obtain
 that the generalized function
\be
 F_n(\psi)(t,x)
&=&
\mu^{2n-1}\int
W^0_2(t,x-x')\psi(x')d x'\cr
&=&\int
e^{i\mu(k)t}e^{-ikx}\mu(k)^{2n-1}\psi^\sim(k)d^3 k
\ee
is equal to zero in ${\cal O}.$
This and
the ``edge of the wedge'' Bogoliubov theorem
imply that this generalized function is
equal to zero.
 Lemma 5.1 is proved.

\medskip
Hence, we construct the transform
$$ P : \Sc_{Re}(\R^4) \to
\Sc(\R^3,\C) \subset H^{1/2}(\R^3,\C),
$$
$$
Pf = \int\mu^{-1}e^{i\mu t}f(t)dt =
\int\mu^{-1}\cos(\mu t) f(t)dt
+ i\int\mu^{-1}\sin(\mu t) f(t)dt.
$$
This gives us a possibility
to consider  operators of localization.
These operators are  connected with
 coherent vectors and we use
 localization operators
 to
 prove the integrability property for Wick kernels.

\medskip
Now we consider operators of localization.

Let $\Sigma$ be an open set in
$\R^3$ and $\Lambda$ be an operator of multiplication
in $\Sc_{Re}(\R^3)$ on the function
$\Lambda(x)$ from $\Sc_{Re}(\R^3).$ This function
is equal to
$1$ on
$$
\{x\in\R^3\;\vert\;
	\mbox{dist}(x,\Sigma)\leq{1\over 2}\delta\}
$$
and to zero on
$$
\{x\in\R^3\;\vert\;
	\mbox{dist}(x,\Sigma)\geq\delta\}.
$$
We introduce the operator of localization
$$
\Lambda(\Sigma,\delta) = \mu^{-1}\Lambda\mu,
$$
defined as a real linear operator on
 $\Sc_{Re}(\R^3)$ and as a complex linear operator
on
$\Sc(\R^3,\C).$ In addition, we define also the
operators of localization
$$
\Lambda_{\C}(\Sigma,\delta)
= R^{-1}\Lambda(\Sigma,\delta)R=
R^{-1}\mu^{-1}\Lambda\mu R =
\pmatrix{\mu^{-1}\Lambda\mu&0\cr 0&\Lambda\cr},
$$
 $$
\Lambda_{Re}(\Sigma,\delta) =
J^{-1}\pmatrix{\Lambda&0\cr 0&\Lambda\cr}J=
\pmatrix{\mu^{-1}\Lambda\mu&0\cr
0&\mu\Lambda\mu^{-1}\cr},
$$
which are correctly defined on
$(\Sc_{Re}(\R^3),\Sc_{Re}(\R^3)).$

We note, that operator of localization
$\Lambda_{\C}(\Sigma,\delta)$ commutes
with the operator of imaginary unit
$J=\pmatrix{0&-\mu^{-1}\cr \mu&0\cr},$
\be
J\Lambda_{\C}(\Sigma,\delta)
&=&
\pmatrix{0&-\mu^{-1}\cr \mu&0\cr}
\pmatrix{\mu^{-1}\Lambda\mu&0\cr
0&\Lambda\cr}
=\pmatrix{0&-\mu^{-1}\Lambda\cr
\Lambda\mu&0\cr}\cr
&=& \pmatrix{\mu^{-1}\Lambda\mu&0
\cr 0&\Lambda\cr}
\pmatrix{0&-\mu^{-1}\cr \mu&0\cr}
= \Lambda_{\C}(\Sigma,\delta)J,
\ee
and the operator of localization
$\Lambda_{Re}(\Sigma,\delta)$  does not
commute with this operator of imaginary unit.
In addition, on the real subspace of the form
$(\Sc_{Re}(\R^3), 0)$ the operators
$\Lambda_{Re}(\Sigma,\delta)$
and $\Lambda_{\C}(\Sigma,\delta)$
coincides.

\medskip
To prove Theorem 5.2 we introduce some notations.

Let  ${\cal A}$  be a set in  $\R^4$ and let
$\Sigma({\cal A})$
 be a causal shadow
\footnote{An other name -- a causal
trace of the set
${\cal A},$ see
\cite{BogLOT87}.}
     of the set
${\cal A}$
on the hyperplane
$t = 0,$
$$
\Sigma({\cal A})
= \{x \in \R^3 \;\vert\;\exists (t,y)
\in {\cal A}\mbox{ and }
 t^2-(x-y)^2\geq 0\}. $$
 In other words, this shadow is a section
of the causal envelope of the set
${\cal A}$ by the hyperplane $t = 0.$
In this case a causal envelope is the causal
envelope generated by the wave equation,
more precisely, by the hyperbolicity of the
wave equation.

\medskip
{\bf Theorem 5.2.}
 {\it
The operator of localization
$\Lambda(\Sigma,\delta)$ is correctly defined
as a linear bounded operator
both
on the space  $H^1(\R^3)$ and on the space
$H^{1/2}(\R^3,\C).$
 Operators of localization  $\Lambda_{Re}(\Sigma,\delta),$
$\Lambda_{\C}(\Sigma,\delta)$ are correctly defined
 as linear bounded operators on the space
$H^1(\R^3) \oplus L_2(\R^3).$

If  ${\cal O}$  is a bounded open set
in $\R^4$ with the causal shadow
$\Sigma({\cal O}) \subset \Sigma,$
then the operator $\Lambda_{Re}(\Sigma,\delta)$
is equal to the identity operator on the subspace
$R^{-1} P \Sc_{Re}({\cal O})$
and $\Lambda_{\C}(\Sigma,\delta)$ is equal to
 the identity operator on the subspace
$R^{-1} P \Sc^{even}_{Re}({\cal O}).$
}

\medskip
{\bf Remarks.}
 1. We use the same notation for operators acting in the
different spaces, but we note that these operators are
consistent with those used previously.

2. We note, that
$P \Sc^{even}_{Re}({\cal O})$ coincides with
$\mbox{ Re}\;P \Sc_{Re}({\cal O})$
(with the choice  $f^\Theta(t,x) = f(-t,x)$
and ${\cal O}^\Theta = {\cal O}).$

3. The operators
$$\Lambda_{Re}(\Sigma,\delta),\quad
\Lambda_{\C}(\Sigma,\delta)$$ are not equal to the identity
operator on $H^1(\R^3) \oplus L_2(\R^3).$ It can be easily shown,
if we take, for instance,
$(\varphi,\pi)_{1}
= (\mu^{-1} \Lambda_1, \mu\Lambda_1)$ for the operator
$\Lambda_{Re}(\Sigma,\delta)$ and
$(\varphi,\pi)_{2} =
(\mu^{-1}\Lambda_1, \Lambda_1)$ for the operator
$\Lambda_{\C}(\Sigma,\delta),$
here $\Lambda_1 \in \Sc_{Re}(\R^3).$
Then
$$
\Lambda_{Re}(\Sigma,\delta)(\varphi,\pi)_{1}=
(\mu^{-1} \Lambda \Lambda_1, \mu \Lambda \Lambda_1)
$$
and
$$
(\Lambda_{\C}(\Sigma,\delta)(\varphi,\pi)_{2}=
(\mu^{-1} \Lambda \Lambda_1,\Lambda \Lambda_1)
$$
so, in particular, for
$\mbox{ supp }\Lambda \cap \mbox{ supp }\Lambda_1
=\emptyset, (\varphi,\pi)_{1} \not = 0,$ while
$$\Lambda_{Re}(\Sigma,\delta)(\varphi,\pi)_{1} = 0,$$
and, analogously,
$(\varphi,\pi)_{2} \not = 0,$  while
$\Lambda_{\C}(\Sigma,\delta)(\varphi,\pi)_{2} = 0.$
This is connected with the fact that
the Fock space is a completion of the space
$\underline\Sc/\underline\Sc_0,$
where $\underline\Sc_0$
is zero radical,
or the completion of
$\underline\Sc({\cal O})/\underline\Sc_0({\cal O}),$
see
\cite{Jos65},
 the Wightman reconstruction theorem and
the Reeh-\-Schlieder theorem, in other words,
the subspace
$P\Sc_{Re}({\cal O})$ has an effectively unit
measure with respect to the
Gaussian cylindrical measure  $d\nu(z)$ only
 ( $d\nu(z)$ is a  measure in the rigorous sense on
 a finite-\-dimensional subspace of
$H^{1/2}(\R^3,\C)$).

4. The similar situation appears when we consider
the operator of the imaginary unit
$J = \pmatrix{0&-\mu^{-1}\cr \mu&0}.$
The operator  $J$ is defined in the usual
sense for the space  $H^1 \oplus L_2$ only. For the
space $H^{1/2} \oplus H^{-1/2}$
it can be defined by corresponding continuation
(by continuity).
The similar situation for an operator of
localization, it is equal to the identity
operator on the causal shadow of the set
 ${\cal O}$ only.
Moreover, for $(\varphi,\pi) = R^{-1} Pf,$ where
$f \in \Sc_{Re}(\R^4),$ i.e. for
$$(\varphi,\pi) = (\int \mu^{-1} \cos\mu t f(t) dt,
\int\sin\mu t f(t) dt)$$
a localization with the help of operator
$R^{-1} \mu^{-1} \Lambda \mu R$ (i.e.
$\mu^{-1} \Lambda \mu$
in the complex space) takes place on the real subspace
only.

On the real subspace
$$
(\varphi,\pi) =
R^{-1} P f, \mbox{ where }\; f \in \Sc^{even}_{Re}(\R^4),
$$
i.e. $(\varphi,\pi) = (\int\mu^{-1} \cos \mu t f(t) dt, \;0),$
we obtain the following relation
\be
\Lambda_{\C}(\Sigma,\delta)(\varphi,\pi)
&=&(\mu^{-1}\Lambda\int\cos\mu t \;f(t) dt, \;0)\cr
&=&(\mu^{-1}\int\cos\mu t \;f(t)dt, \;0)\cr
&=&(\varphi,\pi)
\ee
and the similar relation for
the other choice of a real subspace with corresponding
orthonormal transform
both in the complex
space  $H^{1/2}(\R^3,\C),$ and in the space
$$(H^{1/2}(\R^3) \oplus H^{-1/2}(\R^3),J).$$

\medskip
{\it Proof of Theorem  5.2.}
 Let  $\Sigma \subset \R^3 $ be an open set.
It is clear that for every
 $\Sigma$ there exists a such
open bounded set
${\cal O} \subset \R^4,$ that the causal shadow
$\Sigma({\cal O}) \subset \Sigma.$
Moreover, Lemma  5.1
implies that
$$
\{Pv \;\vert\; v \in \Sc_{Re}({\cal O})\}
$$
is dense in $H^1(\R^3,\C)$ and $H^{1/2}(\R^3,\C)$
and, therefore,
$$
R^{-1} \{Pv \;\vert\; v \in \Sc_{Re}({\cal O})\}
$$
is dense in  $H^1 \oplus L_2$
(and in  $H^{1/2} \oplus H^{-1/2}$).
Theorem  5.4
implies that for $v \in \Sc_{Re}({\cal O})$
$$
\mbox{ supp } \int \cos \mu t \;v(t) dt \subset
\Sigma({\cal O}) \subset \Sigma,
$$
$$
\mbox{ supp } \int \mu^{-1} \sin \mu t \;v(t) dt
\subset \Sigma({\cal O}) \subset \Sigma,
$$
and, thus,
$$
\int(\cos \mu t \;v)(t,x) dt
= \Lambda(x) \int (\cos \mu t \;v)(t,x) dt
$$
$$
\int (\mu^{-1} \sin \mu t \;v)(t,x) dt
= \Lambda (x) \int (\mu^{-1} \sin\mu t \;v)(t,x) dt.  $$
Taking into account
that $$ R^{-1} Pv =
(\int \mu^{-1} \cos \mu t \;v(t) dt,
 \int \sin \mu t \;v)(t) dt), $$
we obtain
$$
\Lambda_{Re}(\Sigma,\delta)R^{-1} Pv = R^{-1} P v.
$$
This equality,
boundedness of the operator
$\Lambda(\Sigma,\delta)$
and Lemma 5.1 imply the assertion
of Theorem 5.2 for the operator
$\Lambda_{Re}(\Sigma,\delta).$

For the operator
$\Lambda_{\C}(\Sigma,\delta)$
and for $v \in \Sc^{even}({\cal O})$ we obtain
$$P v = \mu^{-1} \int \cos\mu t \;v(t) dt,$$
$R^{-1} P v = (\mu^{-1}\int\cos\mu t \;v(t) dt, \;0)$ and
$$
\Lambda_{\C}(\Sigma,\delta)
R^{-1} P v = R^{-1} P v.
$$

Theorem  5.2 is proved.

\medskip
It is easy to see that on $P\Sc_{Re}(\R^4),$
 $\varphi^+\in P\Sc_{Re}(\R^4),$
$$
R\pmatrix{\mu^{-1}\Lambda\mu&0
\cr 0&\mu\Lambda\mu^{-1}\cr}
R^{-1}\varphi^+
= R\pmatrix{\mu^{-1}\Lambda\mu&0
\cr 0&\mu\Lambda\mu^{-1}\cr}
\pmatrix{\mbox{ Re }\varphi^+
\cr\mu\mbox{ Im }\varphi^+\cr}
$$
$$
=R(\mu^{-1}\Lambda\mu
\mbox{ Re }\varphi^+,
 \mu\Lambda \mbox{ Im }\varphi^+)
=
\mu^{-1}\Lambda\mu
\mbox{ Re }\varphi^+
+  i\Lambda\mbox{ Im }\varphi^+.
$$
We rewrite  this and the localization condition
formulated in Theorem 5.2 as a separate corollary.

\medskip
{\bf Corollary  5.3.}
 {\it
For vectors  $$h = (\mu^{-1} \int \cos(\mu t) \,f(t) dt,
\int\sin(\mu t) \,f(t) dt),$$
$f \in \Sc_{Re}({\cal O}),$
 the following equalities fulfill
$$
\Lambda_{Re}(\Sigma,\delta) h =
\pmatrix{\mu^{-1}\Lambda \mu&0
\cr 0&\mu\Lambda\mu^{-1}\cr} h = h.
$$
For vectors
 $h = (\mu^{-1}
\int \cos(\mu t) \,f(t) dt, 0),$
$f^\Theta,$ $f \in \Sc_{Re}({\cal O})$
(i.e. $f \in \Sc^{even}_{Re}({\cal O})$
and the vector  $Pf$ is real)
 the following equalities fulfill
$$
\Lambda_{\C}(\Sigma,\delta) h=
\pmatrix{\mu^{-1}
\Lambda\mu&0\cr 0&\Lambda\cr}
h  = h
=\Lambda_{Re}(\Sigma,\delta) h.
$$
}

\medskip
{\bf Remarks.}
 1. An operator of localization is correctly defined
( in the sense of generalized functions)
on the $H^{1/2}(\R^3) \oplus H^{-1/2}(\R^3)$ also
and the equality of Corollary  5.3
extends
on the space
$H^{1/2}(\R^3\C)$
(by continuity in the sense of generalized
functions).

2. It should be noted that the operator
$\mu^{-1}\Lambda\mu$
in the complex Hilbert space
corresponds
to the complex variant of localization operator.
In this case the operation of complex conjugation
given by time reflection and by standard conjugation
agrees with locality on the real subspace,
or on
$\Sc^{even}_{Re}({\cal O})$
(the other variant,
it agrees with locality on the pure imaginary
 subspace,
or on
$\Sc^{odd}_{Re}({\cal O})$).

\medskip
Now we prove the assertion, that we have used
for the proof of Theorem  5.2.

\medskip
{\bf Theorem 5.4.}
  {\it
Let  $v(t,x)$ be a
twice continuously differentiable function and
$\mbox{ supp } v \subset {\cal O},$ ${\cal O}$
is a bounded open set in
$\R^4,$ then the supports of
$$
v_1(x) = \int(\cos(\mu t) \;v)(t,x) dt
$$
and
$$
v_2(x) =
\int({\sin(\mu t)\over\mu}\;v)(t,x) dt
$$
belong to the compact
$\overline{\Sigma({\cal O}}).$
}

\medskip
{\it Proof of Theorem 5.4.}
The assertion is implied by hyperbolicity of the
free wave equation and its proof is the
same as the proof of finiteness of
propagation velocity of a solution of the free wave
equation, see Reed, Simon
\cite[v.2, \S 13, Theorem X.77]{ReeS75}
and is a consequence of Paley-\-Wiener theorem, see
\cite[v.2, \S 13, Theorem IX.11,
Theorem IX.12 and
also  Theorem IX.13, IX.14]{ReeS75}.
 Theorem  X.77
\cite[v.2, \S 13]{ReeS75}
 implies that
$$
\mbox{ supp }(\cos\mu t \;v)(t,\cdot)
\subset \Sigma(t \times
\mbox{ supp } v(t,\cdot)), $$
$$
\mbox{ supp }({\sin\mu t\over\mu}\;v)(t,\cdot)
 \subset \Sigma(t \times \mbox{ supp }v (t,\cdot)),
$$
and, so, after integration over
$t,$ we have  $$ \mbox{ supp }v_1
\subset \bigcup_{t \in \mbox{ supp }v(\cdot,x)}
\Sigma(t \times\mbox{ supp }v(t,\cdot))
\subset \overline{\Sigma({\cal O}}), $$
$$
\mbox{ supp }v_2
\subset \bigcup_{t \in \mbox{ supp }
v(\cdot,x)}
\Sigma(t \times \mbox{ supp }v(t,\cdot))
\subset \overline{\Sigma({\cal O}}).
$$

The proof of the required assertions
of Theorem X.77
\cite[v.2, \S 13]{ReeS75}
 is the following.

We consider the functions
$$
v_1(x) = (\cos\mu t \;v)(t,x),
$$
$$
v_2(x) = ({\sin\mu t\over\mu} \;v)(t,x).
$$
For each  $t \in
\{t \in \R \;\vert \; \exists x \; (t,x) \in
\mbox{ supp }v\}$
there exists a ball $S^0_r$
with center at zero and with radius
$r,$  such that
$\mbox{ supp }v(t,\cdot) \subset S^0_r,$
moreover, due to compactness of the support of
 $v$ in $R^4,$  the radius of the ball is independent
of  $t.$ Then after the Fourier transform
 (in $x$) we obtain
$$
v^\sim_1(t,k) = \cos t \sqrt{k^2+m^2}v^\sim(t,k),
$$
$$
v^\sim_2(t,k)
= {\sin t \sqrt{k^2+m^2}\over \sqrt{k^2+m^2}}
v^\sim(t,k).
$$
The Paley-\-Wiener theorem (for distributions) implies that
$v^\sim(t,k)$ are entire analytic functions on
$k$ for every  $t$ from the support of $v.$ There exists
a constant  $c$ and an integer number
$N$ (independent of $t$), that
$$
v^\sim(t,k) \leq c(1+|k|^2)^N e^{|Im\,k|\,r}.
$$
Further, $\cos( t \sqrt{k^2+m^2})$
and $(k^2+m^2)^{-1/2} \sin(t \sqrt{t^2+m^2}),$
represented as power series, are entire holomorphic functions
in  $k$ also and they satisfy inequalities
$$
|\cos(t\sqrt{k^2+m^2})|
\leq c_2 e^{|Im\,k|\,|t|},
$$
$$
|(k^2+m^2)^{-1/2} \sin (t\sqrt{k^2+m^2})|
\leq c_3 e^{|Im\,k|\,|t|}.
$$
This means that the functions $v^\sim_1(t,k)$
and  $v^\sim_2(t,k)$ are entire analytical functions
of  $k.$  There exist such constants
 $c_4,$
$c_5,$ that
$$
|v^\sim_1(t,k)|
\leq c_4(1+|k|^2)^N e^{|Im\,k|\,(r+t)},
$$
$$
|v^\sim_2(t,k)|
\leq c_5(1+|k|^2)^N e^{|Im\,k|\,(r+t)}.
$$

Then, due to the
converse
 Paley-\-Wiener theorem  $v_1(t,x)$ and $v_2(t,x)$
have support in
$S^0_{r+t} = \Sigma(t\times S^0_r).$
$\cos(\mu t)$ and $\mu^{-1}\sin(\mu t)$ are
convolution operators. This implies that
if  $v(t,x)$ (for a given $t$) has
compact support $\mbox{supp }v(t,\cdot),$ then for any given
$\varepsilon>0$ one can find a finite number of such balls $S_1,
..., S_M,$ that $$\mbox{ supp }v(t,\cdot) \subset \bigcup^M_{i=1}
S_i \mbox{ and } \bigcup^M_{i=1} S_i \subset \{x \;|\; \mbox{
dist}(x,\mbox{ supp }v(t,\cdot)) \leq \varepsilon\}.  $$ Taking
into account that the solution depends linear on initial data, we
conclude that supports of $v_1(t,x),$ $v_2(t,x)$ belong to the
	set $\cup^M_{i=1} \Sigma(t \times S^0_i).$ This set is
contained in $\Sigma((t + \varepsilon) \times \mbox{ supp }
v(t,\cdot)).$ Since $\varepsilon$  was chosen arbitrary, so
$v_1(t,x)$ and $v_2(t,x)$ have supports in $\Sigma(t \times
\mbox{ supp }v(t,\cdot).$ Namely this assertion has been used in
the previous arguments.  Theorem 5.4 is proved.




\section*{6. Quantum field as an operator--valued
\\ function from
$\Sc^{\alpha\prime}(\R^4).$}

In this Section we consider an operator-\-valued
structure of the fields
$\phi_{in},$ $\phi,$ $\phi_{out}$ and
prove that these fields are operator-\-valued
functions on any  space
$\Sc^\alpha(\R^4),$ $\alpha<6/5.$
This allows us to consider vacuum averages of the fields,
in particular,
to construct as generalized functions
Wightman functions
and matrix elements of the scattering operator.

A space $\Sc^\alpha(\R^4), \alpha>1,$ contains a dense subspace
of functions with compact support in $\R^4.$ This allows us to
consider the locality property
\cite{Osi96c}.

In Section 4 it was shown, that the expression
$$ e^{i(t_1+is_1)H}\phi(t,h) e^{i(t_2+is_2)H},\quad
s_1>0,\quad s_2>0, $$ depends on $t_1+is_1,$ $t_2+is_2$
holomorphically for $s_1>0, s_2>0,$ and is
a Hil\-bert-\-Schmidt operator.
Then we obtained some required estimates. With the help of
these estimates and holomorphy we can consider the boundary
values of these operators and their products. We show that the
fields are operator-\-valued generalized functions, this means
that the fields are operator-\-valued generalized functions on
the space $\Sc^\alpha$ with the appropriate $\alpha,$ namely,
$\alpha<6/5.$

This allows us to consider fields
smoothed with test functions with compact support in
coordinate space and to show that the field
 $\phi(t,x)$
is a strictly local quantum field, see
\cite{Osi96c}.

In this Section we prove the assertions,
that are the direct consequence of the estimates
obtained in Section 4, see Theorem 4.3 and 4.4.

Let $ \phi_\#$
with subscript $\#$ denote either
 the $in$--field  $ \phi_{in},$ or
 the interacting field $\phi,$
 or  the  $out$--field $\phi_{out}.$

\medskip
{\bf Theorem 6.1.}
 {\it
Let  $\chi_1,\chi_2$
be vectors in the Fock space,
let
 $s_1>0,$ $s_2>0.$
An  expression
$$
\langle\chi_1,e^{i(t_1+is_1)H}
\phi_\#(t,h)
e^{i(t_2+is_2)H}\chi_2\rangle,
\eqno(6.1)
$$
depends holomorphically on
$t_1+is_1,$ $t_2+is_2.$
The boundary value for $s_1, s_2 \to 0$
exists  and is a generalized function on
$t_1, t_2$ from the space
$\Sc^{\alpha\prime}$ for any $\alpha<6/5.$

Analogously the expressions
$$
e^{i(t_1+is_1)H}\phi_\#(t,h)
e^{i(t_2+is_2)H}\chi
\eqno(6.2)
$$
$$
e^{i(t_1+is_1)H}\phi_\#(t,h)
e^{i(t_2+is_2)H}
\eqno(6.3)
$$
are, correspondingly, vectors and bounded operators
in the Fock space, depend holomorphically on
 $t_1+is_1,$
$t_2+is_2,$ their boundary values
exist for
$s_1,$ $s_2\to 0$ in the sense of vector-\-valued and
operator-\-valued generalized function on
 $t_1,$ $t_2$ from the space
$\Sc^{\alpha\prime},$ $\alpha<6/5.$
Moreover, for $s_1>0,$ $s_2>0$
the operators (6.3)
are Hil\-bert-\-Schmidt operators.
}

It is possible to formulated
the analogous assertion, that
can be obtained for more general case with the
change  $i(t+is)H$ on
$i(t+is)H+i(x+iy)P.$
We formulate these assertions as the separate theorem.

Let $\Gamma$ be a cone in $\R^4$ and $\Pr \Gamma$
denote the set
$$
\Pr \Gamma = \{(t,x)\in \R^4 | (t,x)\in \Gamma
\; \mbox{ and }
 \; t^2+x^2=1\}.
$$

\medskip
{\bf Theorem 6.2.}
{\it
Let
$\chi_1,\chi_2 \in
 {\Phi o\kappa}, $
$(t_1 + is_1, x_1 + iy_1),
(t_2 + is_2, x_2 + iy_2) \in \R^4 + iV_+.$
  An expression
$$
\langle\chi_1,
e^{i(t_1+is_1)H+i(x+iy_1)P}
\phi_\#(t',x')
e^{i(t_2+is_2)H+i(x_2+iy_2)P}\chi_2\rangle
\eqno(6.4)
$$
depends holomorphically on
$(t_1 + is_1, x_1 + iy_1)
\in \R^4+iV_+,
(t_2 + is_2, x_2 + iy_2) \in \R^4 + iV_+,$
  is $C^\infty$ smooth function on
$t', x'$ and satisfies the estimate
$$ \vert\int\langle\chi_1,
e^{i(t_1+is_1)H+i(x_1+iy_1)P}
\phi_\#(t',x')
e^{i(t_2+is_2) H + i(x_2 + i y_2) P}
\chi_2\rangle h(x')dx'\vert
$$
$$
\leq c_1\Vert\chi_1\Vert_{
\Phi o\kappa}
\Vert
\chi_2\Vert_{
\Phi o\kappa}
\Vert\mu^{-1/2}h\Vert_{L_2(\R^3)}
$$ $$\exp(c(s_1-|y_1|)^{-2\alpha-1}
 +c(s_2-|y_2|)^{-2\alpha-1}),
\eqno(6.5)
$$
where
$\alpha$ is such that
$\Vert\mu^{-\alpha} \Lambda
(\Sigma,\delta)\Vert_{HS(H^{1/2}(\R^3))}
< \infty,$
in particular, any
$\alpha > 2.$

The boundary value for
$(s_1,y_1) \in \Gamma,$ $(s_2,y_2)
\in \Gamma,$ $(s_1, y_1), (s_2, y_2) \to 0$
exists in the sense of generalized function
from the space
$\Sc^{\alpha\prime}(\R^4),$
$\alpha<6/5.$ Here a subcone
$\Gamma \subset V_+$
and $\Pr\Gamma$ belongs to a compact in
$\Pr V_+.$
}

\medskip
{\bf
Remark.}
We note that the boundary value in sense of
operator-valued generalized functions for an expression
$$e^{it_1H-s_1H}\phi e^{it_2H-s_2H}$$
is the generalized function
$e^{it_1H}\phi e^{it_2H}$  (on $t_1,t_2$).
This generalized function takes values in the space of bounded
operators in the Fock space. However, an expression
$e^{itH}\phi e^{-itH}$
is an operator-valued generalized function (on $t$)
and takes values in some space of unbounded operators in the
 Fock space (namely, it acts in some Gelfand triple in the Fock
 space).

 \medskip The following consequences are valid.

\medskip
{\bf Corollary 6.3.}
 {\it
A bounded operator of the
form
$$
e^{i(t_0+is_1)H} \phi_\#(t'_1, h_1)
e^{i(t_1+is_1)H}...
e^{i(t_{n-1}+is_{n-1})H}
\phi_\#(t'_n,h_1)e^{i(t_n+is_n)H}
$$
is correctly defined in the
Fock space and is holomorphic
on  $(t_j + i s_j)$ for $s_j > 0,$
$j=0,...,n,$
as operators in the Fock space. The
boundary value exists as a limit for
$s_j\to 0,$ $j=0,...,n,$ and is an operator-\-valued
generalized function in  $t_0,...,t_n$
over $\Sc^\alpha(\R^{n+1}),$ $\alpha<6/5,$
with values in the space of bounded
operators in the Fock space,
i.e. in  $L(\Phi o \kappa).$

A vector
$$
\prod_{j}
(e^{i(t_j+is_j)H}\phi_\#(t'_j,h_j)
e^{i(t_j+is_j)H})\chi
$$
is correctly defined and
depends holomorphically on
$t_j+is_j,$ $s_j>0,$ $j = 0,...,n,$
as vectors in the Fock space.
The boundary value for
$s_j=0,$ $j=0,...,n,$ exists
as a limit in  $s_j\to 0,$
$j=0,...,n,$ in the sense of vector-\-valued
generalized function on
$\Sc^\alpha(\R^{n+1}),$
$\alpha<6/5.$
}

\medskip
{\bf Corollary 6.4.}
 {\it
A operator of the form
$$
\prod e^{i(t_j+is_j)H+i(x_j+iy_j)P}
\phi(t'_j,x'_j)
e^{i(t_{j+1}+is_{j+1})H+i(x_j+iy_j)P}
$$
is correctly defined as
bounded operators in the Fock space and
 depends holomorphically on
$(t_j+is_j,x_j+iy_j)
\in \R^4+iV_+,$ $j=0,...,n,$
as bounded operators in the Fock space.
 The boundary value for
$(s_j,y_j) = 0,$ $j=0,...,n,$ exists as a limit
in $(s_j,y_j) \in \Gamma,$
$(s_j,y_j)\to 0,$ in a sense of
operator--valued generalized function on
$\Sc^\alpha(\R^{4(n+1)}), \alpha<6/5.$
Here $\Gamma$
is some subcone,
$\Gamma\subset V_+$ and $\Pr\Gamma$
 belongs to a compact in
  $\Pr V_+.$

A vector
$$
\prod_j
(e^{i(t_j+is_j)H+i(x_j+iy_j)P}
\phi_\#(t'_j,x'_j)
e^{i(t_{j+1}+is_{j+1})H+i(x_j+iy_j)P})\chi
$$
is correctly defined and depends holomorphically on
$(t_j+is_j,x_j+iy_j) \in \R^4+iV_+,$
$j = 0,...,n,$
as a vector in the Fock space.
The boundary value for
$(s_j,y_j) = 0,$
$j = 0,...,n,$
exists as a limit in
$(s_j,y_j) \in \Gamma,$
$ (s_j,y_j) \to 0,$ in a sense of vector--valued
function on
$\Sc^\alpha(\R^{4(n+1)}),$ $\alpha<6/5.$
Here $\Gamma$
is some subcone,
$\Gamma\subset V_+$ and $\Pr\Gamma$
 belongs to a compact in
 $\Pr V_+.$
}

\medskip
{\it
Proof of Theorems 6.1, 6.2 and Corollaries 6.3, 6.4.
}
 We note, first, that there are the estimates
$$
\Vert e^{-s_1 H-y_1 P}\phi(t,h)
e^{-s_2 H-y_2 P}\Vert
$$
$$
\leq c_1
\exp(c_0 + c(s^2_1 - y^2_1)^{-\alpha}
+ c(s^2_2-y^2_2)^{-\alpha})\Vert
\mu^{-1/2}h\Vert
$$
$\alpha>5/2$
(take, for instance, $\alpha =
2^{-1}(5+\delta)$ for some
$\delta>0$).

Proof of Theorems  6.1, 6.2
and their consequences is analogous. In particular,
Theorem 6.2
is the corollary of Theorem  4.3 and estimates of Section  4
(or the corollary of estimates of consequences
from Section 4). This corollary is implied by usual assertions
about boundary values.
The required assertions about boundary values
for the spaces
 $\Sc^\alpha,$ $\alpha > 0,$ see, for instance,
 Constantinescu, Thalhaimer
\cite[Theorems 5.1-5.3,6.2]{ConT79}
(these theorems of
 Constantinescu and Thalhaimer
consider the existence of boundary value
 and
  the Four\-ier-\-La\-p\-lace transform).
 Theorems 6.1, 6.2 and Corollaries 6.3, 6.4 are proved.

\medskip
{\bf Remarks.}
1.  We note that the estimates
$\Vert e^{-s_1 H}\phi
e^{-s_2 H}\Vert_{HS}$
appear for $s_1,s_2>0$ only.
Estimates for a  smoothed field
 $\int \phi(t,x)f(t,x)dtdx$ are fulfilled in general for
 the operator norm only and
 in general these estimates are not valid for the
Hil\-bert-\-Schmidt norm.

2. We remark that estimates like $\exp(cs^{-\alpha})$
 for the Fourier--Laplace transform on the boundary
 require a behavior for test functions better than
 $\exp(-c|p|^{\alpha/(\alpha + 1)})$  in momentum
 space, see, for instance,
 \cite[v.2, p. 204]{GelS58}.

\medskip
{\it
Proof of Theorems
3.1, 3.2, 3.3, 3.4, 3.5, 3.6, 3.7.
}
 The assertions
about  properties of Wick kernels and about
 operator-\-valued properties of the field $\phi_\#$
 proved in \cite[Theorem 1.1]{Osi96a},
  \cite[Theorems 3.1-3.4]{Osi96b} and in
 Section 4 (Theorems 4.1-4.4) imply
easily the existence of vacuum averages, in particular, the
existence of Wightman functions as generalized functions, their
properties  as positivity, spectrality, and Poincar{\'e}
invariance.
 Theorems 3.1-3.7  are
 proved.

\medskip
{\it
Proof of Theorem
3.8.
}
We consider the diagonal of Wick kernel of
the quantum field $\phi_{in},$ $\phi,$ $\phi_{out}$
on a coherent vector $e_v,$ $v\in \Sc(\R^3,\C).$
Let $u_{in}$ be the solution of the free linear wave equation
given by the diagonal of the Wick kernel of the free
$in$-coming field $\phi_{in},$
$$
u_{in}(t,x)=
{\langle e_v, \phi_{in}e_v\rangle
\over
\langle e_v, e_v\rangle}.
$$
Then the diagonal of Wick kernel of the
quantum field $\phi$ on the coherent vector
$e_{\varepsilon v}$
is real and on the diagonal the
Wick symbol of the quantum
field is equal to the solution
$u_\varepsilon(t,x)$ of the classical
nonlinear wave equation with $in$-data
which are the initial data of
the solution $\varepsilon u_{in}.$ Namely,
$$
u_\varepsilon(t,x)=
{\langle e_{\varepsilon v},
\phi(t,x) e_{\varepsilon v}\rangle
\over \langle e_{\varepsilon v}, e_{\varepsilon v}\rangle}.
$$
We use the notation
$\langle e_{v_1}, \phi(t,x)  e_{v_2}\rangle $
for the value of the Wick kernel of the quantum
field $\phi $ at point $(t,x)\in \R^4,$ i.e.
 $$\langle e_{v_1}, \phi(t,x) e_{v_2}\rangle
=  \phi (e(e^{it\mu}v_1),e(e^{it\mu}v_2))(x)
$$
This notation was introduced for the formulation of
Theorem 3.8 about nontriviality.

To prove Theorem 3.8 we use the equality proved by
Morawetz and Strauss
\cite{MorS73} for solutions of the nonlinear
equation.
Morawetz and Strauss use the Banach space
$\cal F$ of free solutions. In particular,
$\cal F$ contains the free solutions with initial
data from $\Sc_{Re}(\R^3)\oplus \Sc_{Re}(\R^3).$

Morawetz and Strauss
\cite{MorS73}
proved the following.
Let $u_{in}$ be a free $in$-coming solution, let
 $u_\varepsilon $ and $u_{out,\varepsilon }$ be,
respectively, the solution of
the classical nonlinear equation
and the $out$-going free solution with the
$in$-data given by the solution $\varepsilon u_{in}.$
In particular, $u_1$ has the
$in$-data given by the solution $u_{in}.$
Then (see \cite[Theorem 1]{MorS73})
\begin{eqnarray*}
\lambda
&=&
{1\over 6}\lim_{\varepsilon\to
0}
\varepsilon^{-4}
 (\int u_1(\tau,y)^4 d\tau dy)^{-1}
\\&&
\end{eqnarray*}
$$\int(u_{out,\varepsilon }(t,x)
 \dot u_{out,2\varepsilon }(t,x)
-\dot u_{out,\varepsilon }(t,x)
 u_{out,2\varepsilon }(t,x))
 dx. \eqno(6.6)
$$
For our case of the free solution $u_{in}$
the initial data belong to
 $\Sc_{Re}(\R^3)\oplus \Sc_{Re}(\R^3)$
and $u_\varepsilon (t,x)$ is a nonzero solution.
Since
\begin{eqnarray*}
0&<&\int u_\varepsilon (t,x)^4dtdx
\\&\leq&(\int\sup_{x} u_\varepsilon (t,x)^2
dt)
\sup_t\int u_\varepsilon (t,x)^2 dx
 < \infty, \end{eqnarray*}
so $(\int u_\varepsilon (t,x)^4 dtdx)^{-1}$ is
correctly defined, nonzero and is equal to
 $$
\langle e_{\varepsilon v}, e_{\varepsilon v}
 \rangle^4
 (\int
\langle
e_{\varepsilon v}, \phi(t,x)
e_{\varepsilon v}
\rangle ^4 dtdx )^{-1}.
$$
 As a consequence of asymptotic
 convergence for
 $T\to +\infty$ the expression
$$
{1\over 6}\varepsilon^{-4} \langle e_v, e_v\rangle ^4
 (\int\langle e_v, \phi(\tau,y) e_v\rangle ^4 d\tau dy)^{-1}
$$
$$\times\int (\langle e_{\varepsilon v}, \phi(t+T,x)
e_{\varepsilon v}\rangle
\langle e_{2\varepsilon v},
\partial_t\phi(t+T,x) e_{2\varepsilon v}\rangle
$$ $$- \langle e_{2\varepsilon v}, \phi(t+T,x)
e_{2\varepsilon v}\rangle
\langle e_{\varepsilon v}, \partial_t\phi(t+T,x)
e_{\varepsilon v}\rangle ) dx
\eqno(6.7)
$$
converges and the limit is equal to
$$
{1\over 6}\varepsilon^{-4}
\langle e_v, e_v\rangle ^4
 (\int\langle e_v, \phi(\tau,y) e_v\rangle ^4
 d\tau dy)^{-1}
$$
$$\times \int (\langle e_{\varepsilon v}, \phi_{out}(t,x)
e_{\varepsilon v}\rangle
\langle e_{2\varepsilon v},
\partial_t\phi_{out}(t,x) e_{2\varepsilon v}\rangle
$$ $$- \langle e_{2\varepsilon v}, \phi_{out}(t,x)
e_{2\varepsilon v}\rangle
\langle e_{\varepsilon v}, \partial_t\phi_{out}(t,x)
e_{\varepsilon v}\rangle ) dx.
\eqno(6.8)
$$
 Taking into account that
 $\langle e_{\varepsilon v},
e_{\varepsilon v}
  \rangle  \to 1$ for
$\varepsilon \to 0$
 the expressions
 (6.7) and (6.8) converge for
$\varepsilon \to 0$
 and
 $$\lim_{
\varepsilon \to 0}
 \lim_{T\to +\infty}
 (6.7)=
 \lim_{
\varepsilon \to 0}
 (6.8)
 = \lambda.
\eqno(6.9)$$
 This is the direct consequence of Equality (6.6).
 Theorem 3.8 is proved.

\medskip
{\bf Remarks.} 1. It follows that the coupling
 constant is uniquely
defined by matrix elements of the $out$-field only.

2.  Expression (6.9) can be rewritten  also in the form
\begin{eqnarray*} \lambda&=&
 {1\over 6}\lim_{\varepsilon\to
0}\varepsilon^{-4}
\langle e_v, e_v\rangle
(\int\langle e_v, :\!\phi^4(\tau,y)\!:
e_v\rangle d\tau dy)^{-1} \\&&
\int(\langle e_{\varepsilon v},\phi_{out}(t,x)
 e_{\varepsilon v}\rangle
\langle e_{2\varepsilon v},
 \dot\phi_{out}(t, x)e_{2\varepsilon v})
v\rangle \\&&
-\langle e_{2\varepsilon v},\phi_{out}(t,x)
 e_{2\varepsilon v}\rangle
\langle e_{\varepsilon v},
 \dot\phi_{out}(t, x)e_{\varepsilon v}\rangle) dx.
\end{eqnarray*} Here $:\!\phi^4(t,x)\!:$ is the
4-th degree of Wick normal product of the quantum field
$\phi. $



\section*{Conclusion.}

Thus, we construct
{\it
the relativistic quantum field theory
 with the interaction}
$:\!\phi^4\!:$ {\it in four-\-dimensional space-\-time.}
The construction is nonperturbative, mathematically
rigorous and the constructed theory is nontrivial.
The constructed quantum field $\phi(t,x)$
is an operator-\-valued generalized function on the space
$\Sc^{\alpha}(\R^4)$ for any $\alpha <
6/5,$ i.e. the quantum field is the operator-\-valued
generalized  function of localizable class of Jaffe type.
We
prove that the quantum field satisfies the following Wightman
 axioms:  the condition of positivity, spectrality, and
 Poin\-car{\'e} invariance.  Locality of the interpolating field
 and the condition of asymptotic completeness we consider in the
 next papers.
The proof of locality is based on the consideration of the
 (unitary) representation of the infinite-\-dimensional Lie
 group generated by the classical dynamics, i.e. the Lie group
 generated by the Poisson bracket, the classical canonical
 coordinates and the classical total Hamiltonian,
 compare, for instance, \cite{Pop91}.
  In
 the next paper we prove that for large times the interacting
 quantum field $\phi$ converges asymptotically to the free $in$--
 and $out$--field, moreover, the constructed interacting field
 defines the unique unitary relativistic invariant nontrivial
 scattering operator \cite{Osi96c}.  The considered field is the
 solution of the nonlinear quantum wave equation with the
interaction $\phi^4$ in four-\-dimensional space-\-time, and
nonlinear terms are defined as Wick normal ordering with respect
to the $in$-field.
 However, we think  that this construction is valid in
the case when we define the normal ordering with respect to the
interacting field at time zero.

Here we do not compare our construction and
our approach
(see papers
 \cite{Osi96a,Osi96b},
 \cite{Bae89,Bae92,BaeSZ92},
 \cite{Hei74},
 \cite{MorS72,MorS73},
 \cite{Osi84,Osi94a,Osi94b,Osi94MIT,Osi95b,Osi96c},
 \cite{Pan80-82,Pan81,Pan82},
 \cite{PanS80,PanPSZ91},
 \cite{PedSZ92},
 \cite{Rac75},
 \cite{Seg66,Seg68}) 
 with recent results of Segal et al.
\cite{PedSZ92}
and with the previous results
about triviality
\cite{AlbGH79a,AlbGH79b,Osi79},
 \cite{Aiz82,AizG83,Fro82,Cal88}.
We think that the considered approach is more
appropriate physically (and mathematically!) to the
 infinite-\-dimensional situation than
approximations of papers
\cite{Osi79,AlbGH79a,AlbGH79b},
\cite{Aiz82,AizG83,Fro82,Cal88}.
 We suppose to consider these questions separately.



\medskip
\medskip
\medskip
\medskip
\section*{Acknowledgments.}
\medskip
\medskip

This work is supported partly by RBRF 96-01649.
It is the third paper of the project
 $\phi^4_4 \cap M.$
 I acknowledge
Anatoly Kopilov, Valery Serbo,
 Vasily Serebryakov,
 Ludwig Faddeev, Anatoly Vershik,
 Peter Osipov,   Julia
 and Zinaida
 for the  help.



\section*{Appendix. Bounds on
Hilbert-\-Schmidt norms of some
operators.}

To obtain the bound on the degree $s$ in the uniform estimate
of the integrals over finite-\-dimensional subspaces
we consider the estimate on the Hilbert-\-Schmidt norm
of the operators
 $\mu^{-\alpha}\Lambda\mu^\beta,$ where $\mu$ is the
operator of multiplication on $\mu(p)=(p^2+m^2)^{1/2}$
in the momentum representation
and $\Lambda$ is the operator of
 multiplication on a function
$\Lambda(x)\in \Sc(\R^3)$ in the coordinate representation.
 This estimate defines the spaces of test functions over
which the quantum field will be an
operator--\-valued generalized function.

To obtain the estimate on the Hilbert-\-Schmidt
norms of the operators
$\mu^{-\alpha}\Lambda\mu^\beta$ use the following lemma.

\medskip
\medskip
{\bf Lemma A1.}
{\it
Let $K$  be the operator with the kernel $K(p,q)$ in $L_2(\R^n).$
Then  the Hilbert--\-Schmidt norm of the operator $K$ is equal to
$$
\Vert K\Vert_{HS(L_2)}
=\bigl(\int\vert K(p,q)\vert^2 dpdq\bigr)^{1/2}.  $$
}

\medskip
\medskip
Lemma A1  and inequalities
$$\mu^{2\beta}(q)\leq
c(\mu^{2\beta}(p)+\mu^{2\beta}(p-q)),\quad \beta>0,
$$
imply that
$$
\Vert\mu^{-\alpha}\Lambda\mu^\beta\Vert^2_{HS(L_2)}=
\int\vert\mu^{-\alpha}(p)
\Lambda^\sim(p-q)\mu^\beta(q)\vert^2 dpdq
$$
$$
\leq c(\int\vert\mu^{-\alpha+\beta}(p)
\Lambda^\sim(p-q)\vert^2 dpdq
+\int\vert\mu^{-\alpha}(p)
\Lambda^\sim(p-q)\mu^\beta(p-q)\vert^2 dpdq)
$$
$$
\leq c(\int\mu^{-2(\alpha-\beta)}(p)dp)
\;\sup\int\vert\Lambda^\sim(p-q)\vert^2 dq
$$
$$
+ c\int\mu^{-2\alpha}(p)dp
\;\sup_p\int\vert\Lambda^\sim(p-q)\mu^\beta(p-q)
\vert^2 dq<\infty
$$
for $2(\alpha-\beta)>n$ and $2\alpha>n,$
i.e. for $\alpha>\beta+n/2.$

Thus, for the case of the operator
$$\mu^{-\alpha}\Lambda(\Sigma,\delta)$$
in the Hilbert space $H^{1/2}(\R^3)$ we obtain
the following estimate
$$
\Vert\mu^{-\alpha}
\Lambda(\Sigma,\delta)\Vert_{HS(H^{1/2})}=
\Vert\mu^{-\alpha+1/2}
\Lambda(\Sigma,\delta)\Vert_{HS(L_2)}=
 \Vert\mu^{-\alpha-1/2}\Lambda\mu\Vert_{HS(L_2)}<\infty
$$
for $\alpha+{1\over 2}>1+{3\over 2}.$
Therefore, in the case of three-\-dimensional
space
 $$\Vert\mu^{-\alpha}\Lambda(\Sigma,\delta)
\Vert_{HS(H^{1/2})}<\infty$$
 for $\alpha>2.$

We formulate this estimate as the separate lemma.

\medskip
\medskip
{\bf Lemma A2.}
{\it
Let $\Sigma$  be an open bounded set in $\R^3$
and let $\Lambda(\Sigma,\delta) = \mu^{-1}\Lambda\mu$ be
the operator of localization introduced in
Section 5.
  Then the Hilbert-\-Schmidt norm of the operator
$\mu^{-\alpha}\Lambda(\Sigma,\delta),$ $\alpha>2,$ is finite,
i.e.
$$\Vert\mu^{-\alpha}\Lambda(\Sigma,\delta)\Vert
_{HS(H^{1/2}(\R^3))} <
\infty$$ for $\alpha>2.$
}

\medskip
\medskip
Namely these estimates give us uniform bounds
for the integrals over finite-\-dimensional subspaces
(see Section 4).




\end{document}